\documentclass{aa}
\usepackage{graphics, times}
\usepackage{afterpage}

\def\kms{\hbox{km\,s$^{-1}$}}
\def\mag{\hbox{$^{\rm m}$}}
\def\NH{\hbox{[N\,{\sc ii}]$\lambda$6583\AA/H$_{\alpha}$}}
\def\sk{Sk$-69^\circ279$}
\begin{document}

\msnr{accepted version 4199}

\title{On the Structure and Kinematics of Nebulae around LBVs and 
LBV Candidates in the LMC}

\author {Kerstin Weis \inst{1,2,3,}\thanks{Visiting Astronomer, Cerro Tololo
Inter-American Observatory, National Optical Astronomy Observatories,
operated by the Association of Universities for Research in Astronomy, Inc.,
under contract with the National Science Foundation.}\fnmsep\thanks{Feodor-Lynen
 fellow of the Alexander-von-Humboldt foundation}
}
\offprints{K.\ Weis, Astronomisches Institut, Ruhr-Universit\"at\,Bochum,
Universit\"atsstr.\ 150, 44780 Bochum, \email{kweis@astro.rub.de}}
\institute{Institut f\"ur Theoretische Astrophysik, Universit\"at Heidelberg,
Tiergartenstr. 15, 69121 Heidelberg, Germany
\and
University of Minnesota, Astronomy Department, 116 Church Street SE, 
Minneapolis, MN 55455, USA
\and
Max-Planck-Institut f\"ur Radioastronomie, Auf dem H\"ugel 69, 53121
  Bonn, Germany
}

\date{Received / Accepted}

\authorrunning{K.\ Weis}
\titlerunning{Nebulae around LBVs and LBV\,candidates in the LMC}

\abstract{We present a detailed analysis of the morphology and kinematics of
  nebulae around LBVs and LBV candidates 
in the Large Magellanic Cloud. HST images and
  high-resolution Echelle Spectra were used to determine the size, shape, 
brightness, and expansion velocities of the LBV nebulae around R\,127, R\,143,
  and S\,61. For S Dor, R\,71, R\,99, and R\,84 we discuss the possible
  presence of nebular emission, and derive upper limits for the size and lower
  limits on the expansion velocities of possible nebulae. Including earlier
  results for the LBV candidates S\,119 and \sk\ we find that in general the
  nebulae around LBVs in the LMC are comparable in size to those found in the
  Milky Way. The expansion velocities of the LMC nebulae, however, are 
significantly lower---by about a factor of 3 to 4---than those of 
galactic nebulae of comparable size. Galactic and LMC nebulae show about the
  same diversity of morphologies, but only in the LMC do we find nebulae with
outflow.  Bipolarity---at least to some degree---is found in nebulae in the
  LMC as well as in the Milky Way, and manifests a much more general feature
  among LBV nebulae than previously known.   
}

\maketitle

\section{Introduction}

The most massive stars we know and observe have masses 
above $50$\,M$_{\sun}$  and start as main sequence O stars
with luminosities  of $L\sim 10^{5-6}$\,L$_{\sun}$. They inhabit
the very upper left part of the {\it Hertzsprung-Russell Diagram} (HRD).
After a hot main sequence phase the stars evolve quickly 
towards cooler temperatures and turn into supergiants within few 10$^{6}$
years. Instead of evolving further towards the red, the most massive stars
enter a  phase of very high mass loss  (about 10$^{-4}\,$M$_{\sun}$yr$^{-1}$)
and reverse their evolution back towards hotter effective temperatures.
In this phase the stars are known as 
{\it Luminous Blue Variables (LBVs)\/}. 
The position of the turning point, and therefore of the LBVs (in quiescence)
in the HRD, depends 
on the star's luminosity, and defines the location of the 
{\it Humphreys-Davidson limit\/} (e.g. Humphreys \& Davidson 1979, 1994; 
Langer et al.\ 1994).

Strong stellar winds and possible giant eruptions in the 
LBV phase peel off more and more of the 
stellar envelope, and lead to the  formation of circumstellar 
{\it LBV nebulae\/} (e.g. Nota et al.\ 1995; Weis 2001).
These LBV nebulae are small, typically between 0.2 and about 2\,pc,
and can therefore only be studied in our galaxy and, with the higher resolution
of the {\it Hubble Space Telescope\/} (HST), in 
a few neighboring galaxies such as the {\it Large Magellanic Cloud\/} (LMC). 
Our knowledge about the evolution of the most massive stellar objects in
general, and in the LBV phase in particular, is sparse. 
It is not known what leads to the very
high mass loss rates in these objects and what triggers the giant eruptions. 
These, however, are essential questions to understanding the stellar evolution of massive stars.   
To gain insight into the  LBV phase  and especially 
the formation of LBV nebulae, we studied the 
nebulae around LBVs in the LMC. These LBV nebulae  have formed under 
different conditions, namely at lower metalicity, than those in the Milky Way,
and are therefore of great interest. So far the LBV nebulae in  the
Magellanic Clouds are
the only nebulae, other than the galactic ones, which we are able to resolve
spatially. In this paper we will compare the morphology and 
kinematics of nebulae around LBVs in the LMC
with those in our Galaxy. 

Only a few LBVs are known (roughly 40, several of which are still marked 
as candidates), of which 10 are in our Galaxy 
and 10 in the LMC.
According to Humphreys \& Davidson (1994), the 
following objects in the LMC are LBVs: S Dor, R\,143, R\,127, R\,110, R\,71, 
HDE 269582, and LBV candidates: S\,119, S\,61, R\,84, R\,99. We will add 
Sk$-69\degr279$ to the list of candidates (now numbering 11 
LBVs/LBV-candidates in the LMC) for reasons discussed 
in Weis et al.\ (1997), and strengthened in Weis \& Duschl (2002),
where this object was analyzed in detail.
Among the LBVs in the LMC, R\,143, R\,127, S\,119, S\,61, and
Sk$-69\degr279$ are known to possess a circumstellar nebula. 
In this paper we present a study of LBVs and LBV candidates in the
LMC, concentrating especially on the analysis (morphology and kinematics)  
of the nebulae---if present---using high-resolution 
Echelle spectra and  Hubble 
Space Telescope images. The stars R\,110 and HDE 269582 had to be 
excluded from this work due to a lack of data. 
Spectra and HST images of the LBV candidate
S\,119 have been previously analyzed (Weis et al.\ 2003). 
The results will be added 
and put into context with this work in the final discussion section.  
In the following we always assume 
a distance to the LMC of 50\,kpc (see e.g. Kov{´a}cs 2000; Panagia et
al.\ 1991; Westerlund 1990).

\section{Observations and data reduction}

\subsection{Imaging with the HST}

To study the morphology of the quite small nebulae around LBVs, 
images from the HST are especially useful.
The following objects have been observed with the HST (see Table
\ref{tab:hstdata}) and analyzed in this paper:
R\,127, R\,143, S\,61, R\,71, S Dor, R\,99. For S\,119, see
Weis et al.\ (2003). For R\,110, R\,84, HDE 269582, 
and Sk$-69\degr279$, no HST observations exist so far.

All objects were observed with the {\it Wide Field Planetary Camera 2
(WFPC2\/)}. 
The  F656N filter was selected and mimics quite well an H$_{\alpha}$ filter,
since the radial velocities of the LMC stars is roughly 250\,\kms, and
therefore H$_{\alpha}$ is within the maximum throughput of the filter. 
All available data were retrieved from the STScI data archive and 
reduced with the usual routines in
STSDAS/IRAF. All (typically there were 4 images) 
longer exposures (500\,s) of one object were 
combined and cosmic-ray cleaned. 
Information about the HST datasets are compiled in 
Table \ref{tab:hstdata}.
The images were not rotated, to ensure the full resolution was maintained.
The celestial directions are therefore indicated in the 
images. The roll angle of the HST images are also given in the last
column of Table \ref{tab:hstdata}. 
All HST images were taken with the relevant stars centered on the PC chip,
which has a sampling of 0.0455\arcsec/pixel. Most figures shown here  of the
stars and nebulae contain only this PC section.
In the case of R\,143 and R\,99, nevertheless, the full mosaiked images are
shown to discuss the stars and nebulae in context with the surrounding ISM.
The sampling of those images is lower with 0.0996\arcsec/pixel. 
For S Dor, R\,71, and R\,99 we subtracted the  Point Spread Function (PSF), 
which was generated with the Tiny Tim (Krist 1995)
software for the corresponding positions and filters of each star. 
In all cases we tended to slightly oversubtract to make sure that possible
residuals are real.

\begin{table*}
\caption[]{\small Compilation of the parameters of the datasets}
\begin{flushleft}
\begin{tabular}{ccccccc}
\hline
 star & HST image & HST Prog. & Echelle Spectra & NTT image & CTIO image & HST
 roll angle \\
\hline
  R\,127  & F656N & 6540 & 5  & - & - & 147.34\degr \\
  R\,143  & F656N & 6540 & 6 & - & H$_{\alpha}$   & 145.97\degr \\
  S\,61  &  F656N & 6540 & 3 & - & - & 145.01\degr \\  
  S Dor & F656N & 6540 & 1 & - &  H$_{\alpha}$ & 17.41\degr \\  
  R\,71 & F656N  & 6540 & 1 & - & - & 21.23\degr \\  
  R\,99 & F656N & 6540 & 1 & H$_{\alpha}$/EMMI & - & 148.58\degr \\
  R\,84 & - & - & 1 & H$_{\alpha}$/EMMI and SUSI & - & -\\
\hline
  S\,119 & F656N & 6540 &  5 & - & - & 148.5\degr \\
  \sk & - & - & 5 & - & H$_{\alpha}$ & - \\
\hline
\end {tabular}
\end{flushleft}
\label{tab:hstdata}
\end{table*}

\subsection{Imaging with the 0.9\,m-CTIO-telescope \label{sect:image09}}

In addition to retrieving the HST images, we also obtained ground-based 
observations made with the 0.9\,m-telescope at the  Cerro Tololo 
Inter-American Observatory. These images were taken with an 
H$_{\alpha}$ filter, which contained the [N\,{\sc ii}]-lines at 
6548\,\AA\ and 6583\,\AA. The filter was centered on 6563\,\AA, and the 
FWHM was about 75\,\AA.
The images were calibrated with bias and sky flatfield frames of the
corresponding nights. The seeing was about 1\farcs4, and the nights were 
photometric. Exposure times ranged between 600\,s and 900\,s for a single 
image. The scale for all images is 0\farcs397 per pixel. All images from
the 0.9\,m-telescope are displayed with north to the top and east to the left. 

\subsection{Imaging with the ESO NTT}

We used the ST-ECF/ESO archive to obtain 
H$_{\alpha}$ or [N\,{\sc ii}] images of the
LBVs, for which no HST images  are  available, or additional information
on the larger environment would be useful for interpreting our long-slit
Echelle data.

The first archival data set was observed with the red arm of the EMMI
multimode focal reducer instrument mounted at the ESO NTT 3.5\,m-telescope.
The filter used in these observations was an H$_{\alpha}$ filter (ESO \#596)
with a central wavelength of 6547\AA\ and a FWHM of 73\,\AA.
The EMMI red arm gives a pixel scale of $0\farcs27$ with the 24\,$\mu$m pixel
CCD (ESO \#36).  Due to limitations from the
optical set-up, the usable field size was $9\farcm2\,\times\,8\farcm6$.
Seeing during the observations was $0\farcs8$, and the expose time was
100\,s for both fields.  We reduced the data in the standard manner using
IRAF. Cosmic rays were corrected on the science frames using the
LA-COSMIC IRAF scripts, which perform cosmic ray detection using a
Laplace filter technique (van Dokkum 2001).

The second archival data set was observed with the SUSI imager, also at the
ESO NTT. A chronographic unit was inserted, which resulted in a round
field with an occulting bar running across the field. The set-up and
reduction of the data are the same as already reported in e.g. Weis (2000).
Integration time of the R\,84 H$_{\alpha}$ image was 1000\,s, and 
the seeing was $0\farcs75$. The cosmic ray hits were corrected using 
the LA-COSMIC IRAF scripts.

As for the CTIO 0.9\,m data, all ESO NTT images are 
displayed with north to the top and east to the left.

\subsection{Long-slit Echelle spectroscopy\label{sect:echelle}}

High-resolution long-slit Echelle observations were made 
with the Echelle spectrograph on the 4\,m-telescope at the Cerro Tololo
Inter-American Observatory in order to study the kinematics of 
the nebulae around R\,127, R\,143 and S\,61. Spectra of S Dor, R\,71, R\,84,
and R\,99 were also taken, in order to search for nebular emission possibly 
connected to the star. 
All observations were made using the same configuration. 
In this set-up for the Echelle spectrograph  we replaced the cross-disperser 
by a flat mirror and inserted a post-slit H$_\alpha$ filter (6563/75\,\AA) 
for order selection. The long-slit, which resulted from this 
configuration, was vignetted to a length 
of  $\sim4\arcmin$. The spectral region that was covered contained
the  H$_{\alpha}$ line as well as two [N\,{\sc ii}] lines at 6548\,\AA\ and 
6583\,\AA. 
We used the  79\,l\,mm$^{-1}$ Echelle grating with a slit-width 
of 150\,$\mu$m, corresponding to an  instrumental FWHM
at the H$_\alpha$ line of about 8\,km\,s$^{-1}$.
The data were recorded with the long focus red camera and a 
$2048\,\times\,2048$ pixel  CCD, with a pixel size of
0.08\,\AA\,pixel$^{-1}$ along the dispersion and 0$\farcs$26\,pixel$^{-1}$
on the spatial axis. The seeing was between 1-2\arcsec\ during the 
observations, and the weather was not photometric.
Thorium-Argon comparison lamp frames were taken for wavelength calibration
and geometric distortion correction.
Telluric lines visible in the spectra were used to improve the absolute
wavelength calibration, the accuracy of which is estimated to
be 0.04\,\AA\ (2\,\kms) or better.

For each object, the {\it position angle\/} (PA) of the 
slit was chosen to match the
symmetry of the nebula, as far as the morphology was previously known.
Therefore, the PAs are different for each 
object and they are described and shown in images
in the corresponding sections. 
The Slit naming was as follows: The Slit is named according to the 
position angle, e.g. Slit PA=222  is a slit with a position angle of
222\,\degr. If several Slits are taken the names indicate the parallel 
offsets of the slits from the center, e.g., Slit 3N is a Slit 3\arcsec\ north 
(or S for south) of the Slit observed at the stars center. 
Generally the naming convention becomes 
obvious from figures in each section which show the position and names of each
Slit for each object.

The data are presented as echellograms and
as measured {\it position velocity diagrams\/} ($pv$-diagrams). 
All echellograms displayed extend 65\,\AA\ along
the spectral axis, which is centered 
on H$_{\alpha}$ (in rest); in spatial direction the echellograms are
centered on the projected position of the central star onto the slit and
are in most cases 1\arcmin\ long.
Some echellograms have an insert (20\arcsec\ high, 5\,\AA\ wide) at
the upper left, which illustrates the emission of the stronger 
[N\,{\sc ii}] line (6583\,\AA) again, at different brightness levels
to show structures with different surface brightness.
For R\,99 and R\,84, the insert includes the total spectrum 
of the star and not only
the [N\,{\sc ii}], to better illustrate the spectral shape at the star's
position in contrast to the emission from the larger surroundings.
The spectra of S Dor, R\,99, and R\,84 are also depicted in their full 
vignetted length of 4\arcmin.  
The $pv$-diagram of each slit 
was obtained by measuring the brighter [N\,{\sc ii}] line at 
6583\,\AA. The zero 
position in the $pv$-diagram corresponds to the projected position of 
the central star onto the slit. 
All $pv$-diagrams are generated by binning 4 pixels (1.04\arcsec)
along the spatial axis. All velocities
are measured in the heliocentric system and show a redshift caused by the 
radial velocity of the LMC.

\section{The morphology and kinematics of individual objects}

\subsection{R\,127}\label{section:r127}
\subsubsection{Previous work}

Henize (1955) first recognized that the star R\,127 
(HDE 269858), in the Large Magellanic Cloud, shows an emission line spectrum.
Walborn (1977, 1982) classified the star later as  Ofpe/WN9. 
The S\,Dor typ variability (spectrum changed from Ofpe/WN9 to an early A, 
late B) of R\,127 led to its classification as an
LBV (Stahl et al. 1983).
For the stellar parameters of R\,127 in the 
maximum of  the S Dor phase, they derived  $M_{\rm bol}= -10.6$\mag,  
$T_{\rm eff} = 16\,000\,$K, and
R$_* = 150\,{\rm R}_{\sun}$, a wind velocity of 
$v_{\rm wind} = 110$\,\kms\ and mass loss rate of 
$\dot{M} = 6\,10^{-5}\,$M$_{\sun}\,{\rm yr}^{-1}$. The star's mass 
estimate is about  60\,M$_{\sun}$.
A good compilation of the photometry
and light curve of R\,127 can be found in van Genderen et al.\ 
(1997a).

Already early low resolution spectra  showed typical nebular lines,
which hinted at the existence of a circumstellar 
nebula (Walborn 1982) and a line split found by Stahl \& Wolf (1986b)
showed the nebula's expansion with roughly $30-40$\,\kms. 
Stahl (1985, 1987)
showed, with direct imaging, that the point spread function of the star 
was slightly extended, and added evidence for a nebula.  
Stahl's estimates for the size of the nebula was 
3\farcs5\,$\times$\,4\farcs5  or 0.8\,$\times$\,1.1\,pc.
This measurement was supported by long-slit observations
(Appenzeller et al.\ 1987) which yield an expansion velocity of 
$v_{\rm exp} = 28\,$\kms, a diameter of 4\arcsec, and
indicated deviation of the nebula's shape from spherical symmetry. 
Polarimetric observations (Schulte-Ladbeck et al.\ 1993) 
revealed also that the stellar wind is asymmetric. Clampin et al.\ (1993) 
were successful in resolving the nebula with the use of the John Hopkins
Adaptive Optics Coronograph.
Their H$_{\alpha}$+[N\,{\sc ii}] image with a resolution of 0\farcs7 
showed a much larger nebula, extending 1.9\,$\times$\,2.2\,pc, and revealed a 
diamond shape.
In a more recent kinematic study, Smith et al.\ (1998) favored a 
model with two expanding
shells  around R\,127: one inner shell (about 0.6\,pc from the star) expanding
with $v_{\rm exp} = 29\,$\kms\ and an outer shell expanding 
with $v_{\rm exp} = 25\,$\kms. Their abundance 
analysis made with an HST-FOS spectrum at the brighter eastern part 
of the nebula
lead to the following parameters of the nebula: log N/H = 8.05, log O/H = 8.10,
$n_{\rm e} \sim 720$ cm$^{-3}$, and $T_{\rm e} = 6420$\,K.

\subsubsection{The morphology from the HST images}

\begin{figure}
{\resizebox{\hsize}{!}{\includegraphics{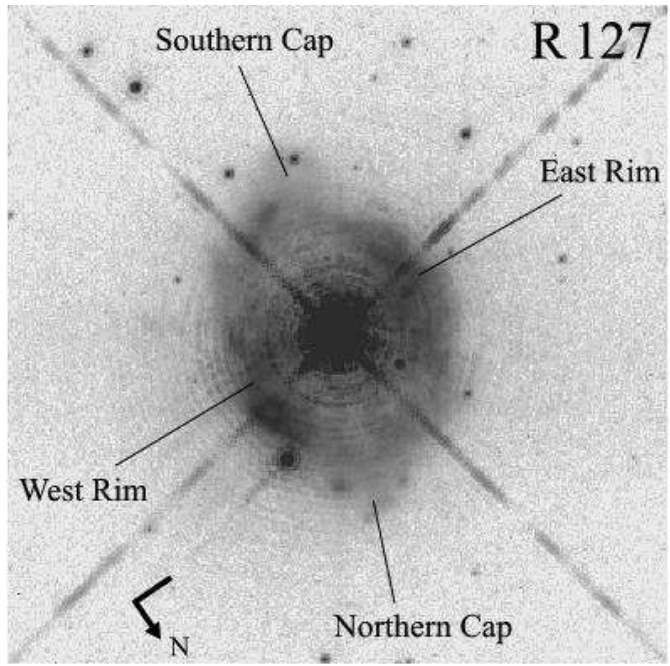}}} \caption{
This figure shows the HST image of R\,127 taken with the F656N
(H$_{\alpha}$) filter. The field of view
is about 15\arcsec\,$\times$\,15\arcsec. A north-east 
vector indicates the celestial orientation. Faint periodic rings 
result from the stars point-spread function.
} \label{fig:r127hst} 
\end{figure}

With the high-resolution HST image, a  detailed
study of the morphology of the nebula around R\,127 is possible.
Fig. \ref{fig:r127hst} shows an F656N filter (H$_{\alpha}$) image of the 
nebula. While the central region of the nebula
resembles  a nearly spherical structure with a diameter of 5\farcs4 
(1.31\,pc), much fainter emission roughly 
north and south adds an elongated shape to the appearance
of the nebula, henceforth called the {\it Northern\/} and
{\it Southern Caps\/}. The Northern and Southern Caps extend 1\farcs27 
(0.31\,pc) and 1\farcs72 (0.42\,pc), respectively, 
beyond the central body.  
The central shell consists of two brighter rims, one at the east, 
designated the {\it East Rim\/}, and one at the (north)-west side, 
the {\it West Rim\/}.
The surface brightness is highest at a knot-like structure
in the West Rim, quite close in projection to a star. The
surface brightness of the nebula is highly variable. 
Beside obvious, distinct low and high
surface brightness areas (Caps and Rims), knots and filaments of higher
surface brightness can be found across the entire nebula. 
These knots reveal sizes  of at least 0\farcs1  or 0.03\,pc for the smallest
resolved structures, and up to 1\farcs26 (0.3\,pc) for the 
largest (and brightest) knot visible in the West Rim. 
The faintest structures in the nebula are the Caps with 
$2.8\,10^{-14}$ ergs\,cm$^{-2}$\,s$^{-1}{\rm arcsec}^{-2}$  
for the Southern Cap
and $3.7\,10^{-14}$ ergs\,cm$^{-2}$\,s$^{-1}{\rm arcsec}^{-2}$ for the 
Northern. Typical regions in the nebula reach a surface brightness 
of $1.2\,10^{-13}$ ergs\,cm$^{-2}$\,s$^{-1}{\rm arcsec}^{-2}$, the East Rim 
is about a factor of 2.5 higher, and the high surface 
brightness knot in the West Rim is about 4.5 times brighter than this. 

\begin{figure}
{\resizebox{\hsize}{!}{\includegraphics{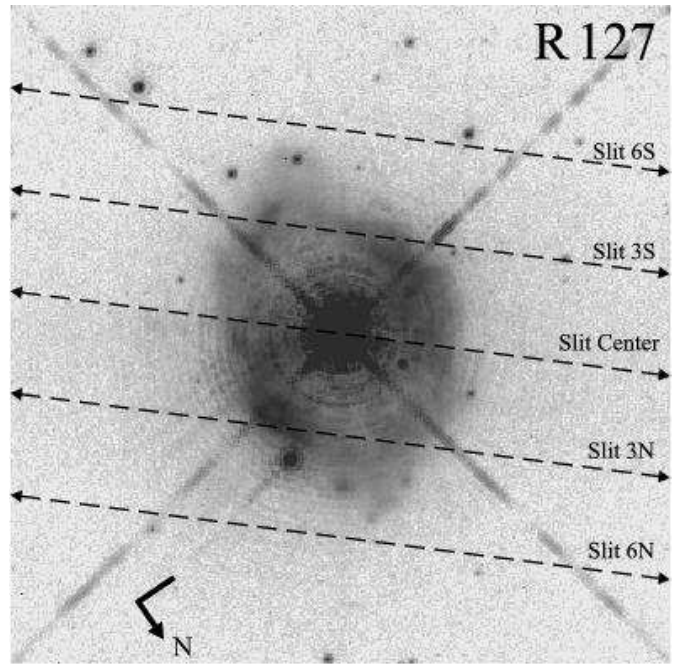}}} 
\caption{ 
Here, the same image of R\,127 as in Fig. \ref{fig:r127hst} is shown, 
and the positions of the slits (PA=230\degr) are marked. 
Note that the slit width was about 1\arcsec.  
} \label{fig:r127slits} 
\end{figure}

\subsubsection{The Kinematics}

\begin{figure*}
\begin{center}
{\resizebox{\hsize}{!}{\includegraphics{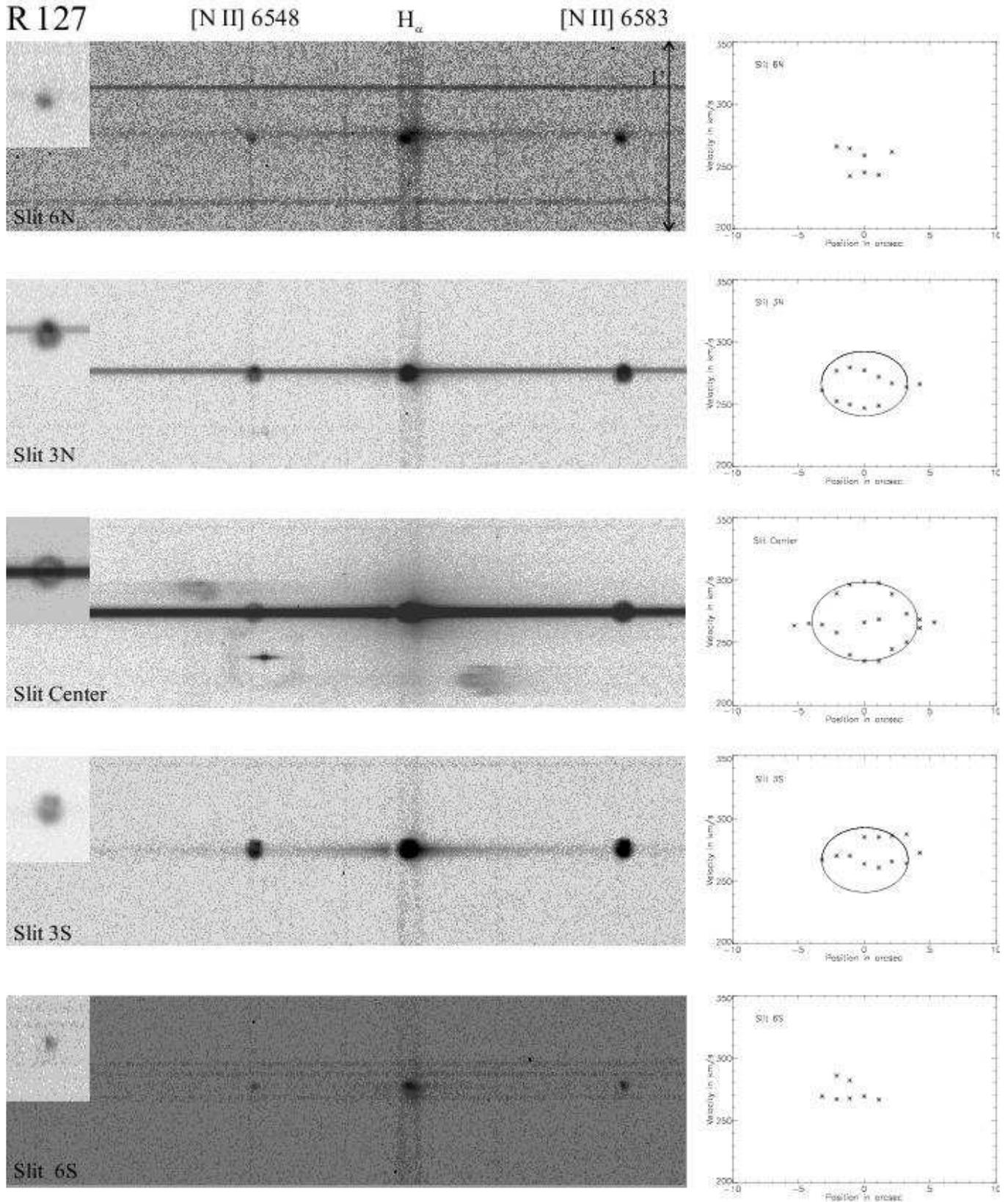}}} 
\end{center}
\caption{Echellograms (left column) and corresponding position-velocity 
diagrams (right column) of our slits for R\,127. 
Measurements are with respect to the 
heliocentric system. South-west (positive positions) is up, 
north-east (negative positions) down. 
In the $pv$ diagrams of Slit 3N, Center and 3S models of spherically expanding
Doppler ellipses are overplotted. 
In Slit Center several
ghost images are present.} \label{fig:echeller127} 
\end{figure*}

We observed 5 spectra of R\,127 with a position angle 
(PA) of PA=230\degr.
The position of the spectra and naming can be found in 
Fig. \ref{fig:r127slits}. The corresponding echellograms and
position-velocity diagrams of each 
slit for R\,127 are shown in Fig. \ref{fig:echeller127}.
For R\,127 the top of each echellogram points 
towards  the south-west and negative offsets in the $pv$-diagram
are to the north-east, positive to the south-west (see also
Fig. \ref{fig:r127slits}).

The radial velocity  of the nebula's
center of expansion lies at about $v_{\rm rad} = 267\,\kms$ and   
is consistent with the stars' velocity 
and H\,{\sc i} measurements of the LMC
(e.g. Rohlfs et al.\ 1984).
With an instrumental  FWHM of roughly 8\,\kms, the expansion ellipse 
of the nebula was clearly resolved. At Slit Center, which crossed the central
star, the expansion velocity was 31.5\,\kms. The expansion is much 
slower in Slit 3N and Slit 3S. In Slit 3N the maximum expansion velocity is 
$v_{\rm exp} = 15$\,\kms, found at position $-0\farcs5$, while the maximum
expansion found in Slit 3S is  
$v_{\rm exp} = 12.3$\,\kms\ at position 1\farcs1. 
In both slit positions the Doppler ellipse is slightly deformed, indicating a
deviation from  spherical expansion. If we determine the spatial center of the 
ellipse (point of half the width) in the $pv$ plots, it would be 
at $-0\farcs3$ in Slit 3N and at +1\farcs5 in Slit 3S. Both ellipses are
much smaller than in the central slit position, and their centers are shifted
against each other.

Futher out, no clear expansion ellipse was 
detected. Nevertheless, the measured velocity values were scattered around 
two components---in the case of Slit 6N, at 243\,\kms and 260\,\kms, and for 
Slit 6S, at 267\,\kms\ and 283\,\kms. The emission detected in Slit 6N can be
identified with the Northern Cap, that in Slit 6S with the Southern. 

Note that at the redshifted side of the Doppler ellipse in Slit 3N a brighter
knot is visible (above the stellar continuum, see insert in Fig. 
\ref{fig:echeller127}), which is
identified with the brighter knot found in the West Rim.

In Slit Center an additional velocity component appears which has a velocity
of 267\,\kms, similar to the radial velocity of the
star (Stahl \& Wolf 1986a) and identical with the velocity of the center of
expansion. 
This component is only visible at the star's position and does not extend
further into the nebula. The large spatial size of this component 
in the $pv$-diagram of Slit Center is only due to the binning of the 
data points.

The \NH\ ratio of the nebula around R\,127, as measured in the spectra 
reaches $0.6 \pm 0.05$, this is consistent with earlier 
measurements (see for example Smith et al.\ 1998).
The Northern and Southern Caps show the same value.
 
\begin{figure}
{\resizebox{\hsize}{!}{\includegraphics{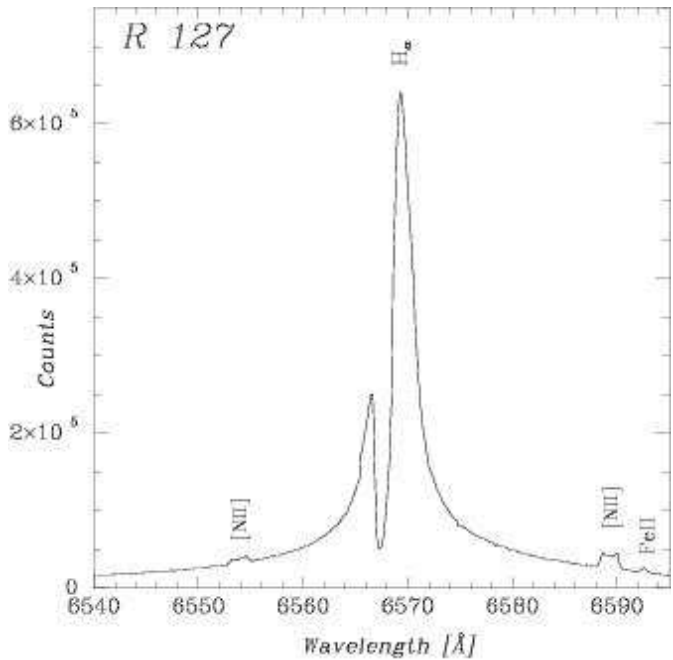}}} 
\caption{ 
This plot shows the stellare spectrum of R\,127 as extracted from the spectrum
in Slit Center. The nebular emission in the line of sight to the stars is still
superimposed. Even though the H$_{\alpha}$ line looks like a P Cygni
type profile, the blueshifted wing shows additional emission, which is not
seen in classical P Cygni lines. The spectrum also shows---barely
recognizable---the three peaks in both [N\,{\sc ii}] lines. 
Next to the [N\,{\sc ii}] line at 6583\,\AA\ the Fe\,{\sc ii} at 6587\,\AA\ is
detected.} \label{fig:r127spectrum} 
\end{figure}

A spectrum of the star (within our limited spectral range)
and the superimposed emission of the nebula is shown
in Fig. \ref{fig:r127spectrum}.  
The tripel peak in the [N\,{\sc ii}] emission lines is faintly
visible.

\subsubsection{Discussion}\label{section:dis127}

Combining the morphological information from the 
HST images (Fig. \ref{fig:r127hst}), 
with the 
kinematics (Doppler shifts) from the spectra, it becomes clear that the 
nebula around R\,127 is not simply spherical. The nebula 
consists  of an inner, nearly spherical part, 
which further out is more and more elongated. 
This elongations is most prominent in the north and south where the 
Caps are attached.
The kinematics of the nebula hint that even the apparently
sherical shell in the center is not expanding as expected (assuming a constant
expansion velocity). 
The deformed expansion ellipses
in Slit 3N and Slit 3S indicates a deviation from spherical expansion. 
This becomes obvious by comparing the data (asterisks) with the model of an 
expansion ellipse with spherical expansion (line) in the $pv$ plots.
In both Slit 3N and Slit 3S the data are shifted with respect to 
the sphecial model. 
If the nebula would be spherical the center of expansion 
should be the same in all slits and lie at about 267\,\kms.  
Therefore it is concluded that the nebula 
is not expanding with sphercial symmetry in the north and south.
More precisely, the southern part of the nebula is 
more redshifed, the northern more blueshifted than the nebula's center.
The global expansion is bi-directional.  
This bi-directional expansion is supported by the image of the nebula,
which shows an elogated bipolar shape, most obvious in the Caps.

In addition to the 
expanding shell, [N\,{\sc ii}] emission is also  present 
with the radial velocity of the star, but  only at 
the star's position. 
This emission may result either from a second shell which is
extremely close to the star, or represents a 
knot-like structure which 
appears projected onto the star and moves more slowly than the shell.
The emission might also have its origin in a circumstellar 
disk close to the star which was proposed by Schulte-Ladbeck 
et al.\ (1993). 
This  third velocity
component (two result from the expansion ellipse of the shell) 
is clearly separated 
from the first two components 
of the expansion ellipse. It is a single peaked 
emission line with a  FWHM=23.6\,\kms (corrected for the
instrumental FWHM). 
A determination from  the HST images of a second shell is impossible
because of the bright central star. It also shows some 
bleeding which effects the longer exposures and makes it 
almost impossible to identify any features within
1\arcsec\ of the star.
A possible second---inner---shell around R\,127 was already 
suggested by Smith et
al. (1998), and they quote an expansion velocity of 29\,\kms. This result 
is not supported by the new measurements presented here and might result from 
a mixing of the three components. 
     
Most likely the third [N\,{\sc ii}] emission line structure
is indeed a knot within the nebula or closer to the star. 
This scenario would best explain the 
similarity of the radial velocity of this emission line structure with
the star's radial velocity.

\subsection{R\,143}

\subsubsection{Previous works}

Even though the  central cluster of 30\,Doradus is home to 
many young and massive stars, only one LBV is known in this region, 
the star R\,143 or HD 269929. 
This star is located roughly 2\farcm2 south of R\,136, the dense cluster and 
core of 30\,Dor. Based on a comparison of different photometric 
studies with a timespan 
of about 40 years (Parker 1992) and spectral analysis,  R\,143 was 
classified as LBV by Parker et al.\ (1993). During this time, the 
star changed from a relatively cool star (late F-type) to the hot part of the 
HRD (early B, perhaphs O9.5) and then cooled again to late B. 
Parker et al.\ (1993) estimated 
$M_{\rm bol} \sim -10$\mag\ and $M_{\rm ZAMS} \sim 60\,$M$_{\sun}$.   

\begin{figure}
{\resizebox{\hsize}{!}{\includegraphics{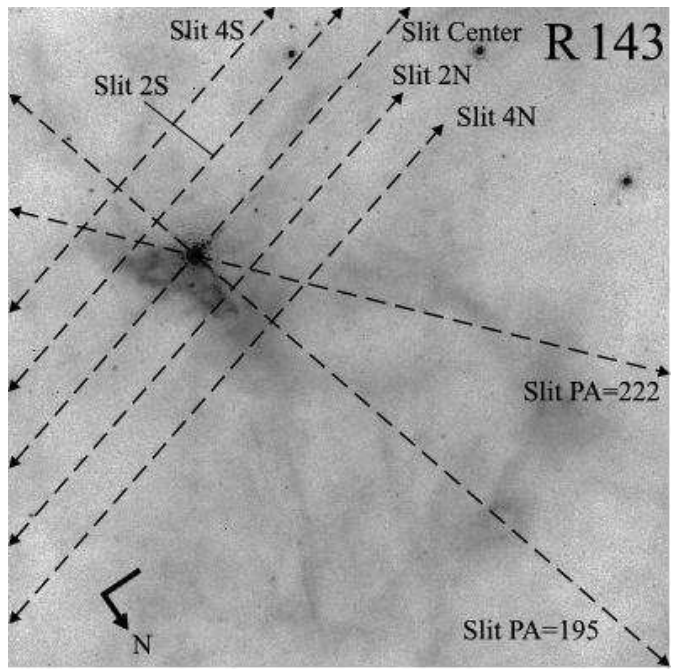}}} 
\caption{ 
A larger section (30\arcsec\,$\times$\,30\arcsec) of the HST image  
showing the location of our slits taken from the nebula around R\,143,
including Feasts' filaments.The HST image clearly shows  that these filaments 
bend towards the star but are not connected to the much 
smaller  LBV nebula. Re-determining the sizes from the HST image, 
the filaments extend to 16\farcs28\ (3.9\,pc) and 16\arcsec (3.8\,pc)
north-east and north of the star, respectively.  
} \label{fig:r143slits} 
\end{figure}

The star was identified by Feast et al.\ (1960) 
and classified as an F7 {\sc i}a star.
Feast (1961) also identified at least four curved filaments 
close to the star, two of which could belong to a nebula around R\,143
and are 15\arcsec\ long (3.5\,pc, see Figs.\ \ref{fig:r14330dor} and
\ref{fig:r143slits}). 
Closer to the star, an elliptical 
structure was found  slightly to the west of the central star.
The identification of the filaments as part of a circumstellar nebula
is hindered by many filaments and knots which belong to 
the 30\,Dor H\,{\sc ii} region---a clear classification of a 
circumstellar nebula 
from morphology alone is not possible.
The first high quality images of the nebulosity around R\,143
were obtained  by Clampin et al.\ (unpublished, as quoted in 
Nota et al.\ 1995). These images show 
stronger [N\,{\sc ii}] emission close to the star.  
Smith et al.\ (1998) analyzed the nebula around R\,143 and found that the long
filaments show abundances similar to those of the 30\,Dor H\,{\sc ii} region.
Therefore, it was concluded that the filaments are not part of an LBV nebula,
which should show a higher nitrogen content due to the CNO processed material.
The filaments also move with a nearly constant radial velocity which is of the
order of the velocities found in the 30\,Dor complex.
However, R\,143  is surrounded by a much smaller high surface brightness 
LBV nebula.
Spectra (Smith et al.\ 1998) very close to the star 
revealed nitrogen enhanced material. This area
coincides with the stronger [N\,{\sc ii}] emission region seen by Nota et al.\
(1995, Clampins images, see above). About 1\farcs8 north of the star ``a 
bona fide LBV nebula'' is present. Smith et al.\ (1998) derive 
for the nebula a 
$T_{\rm e} = 12\,200\,$K, an average  $n_{\rm e} < 100\,$cm$^{-3}$, and 
 $n_{\rm e} = 1000\,$cm$^{-3}$ in the inner 5\arcsec.   
Their \NH\ ratio of the nebula varies according to the distance 
from the star and reaches a
maximum of 0.8 at 1\farcs8 north of the star.
This nebula  is elongated in the north-south 
direction with a diameter of about 5\farcs2
(1.3\,pc). Their spectra  showed  a 
blueshifted motion with a velocity difference of 130\,\kms. They 
concluded that
the LBV nebula around R\,143 is much smaller, and 
the filaments (Feast 1961) are part of the 30\,Dor complex and most likely 
not associated with R\,143.

\subsubsection{The Morphology from HST images}

\begin{figure*}
\resizebox{\hsize}{!}{\includegraphics{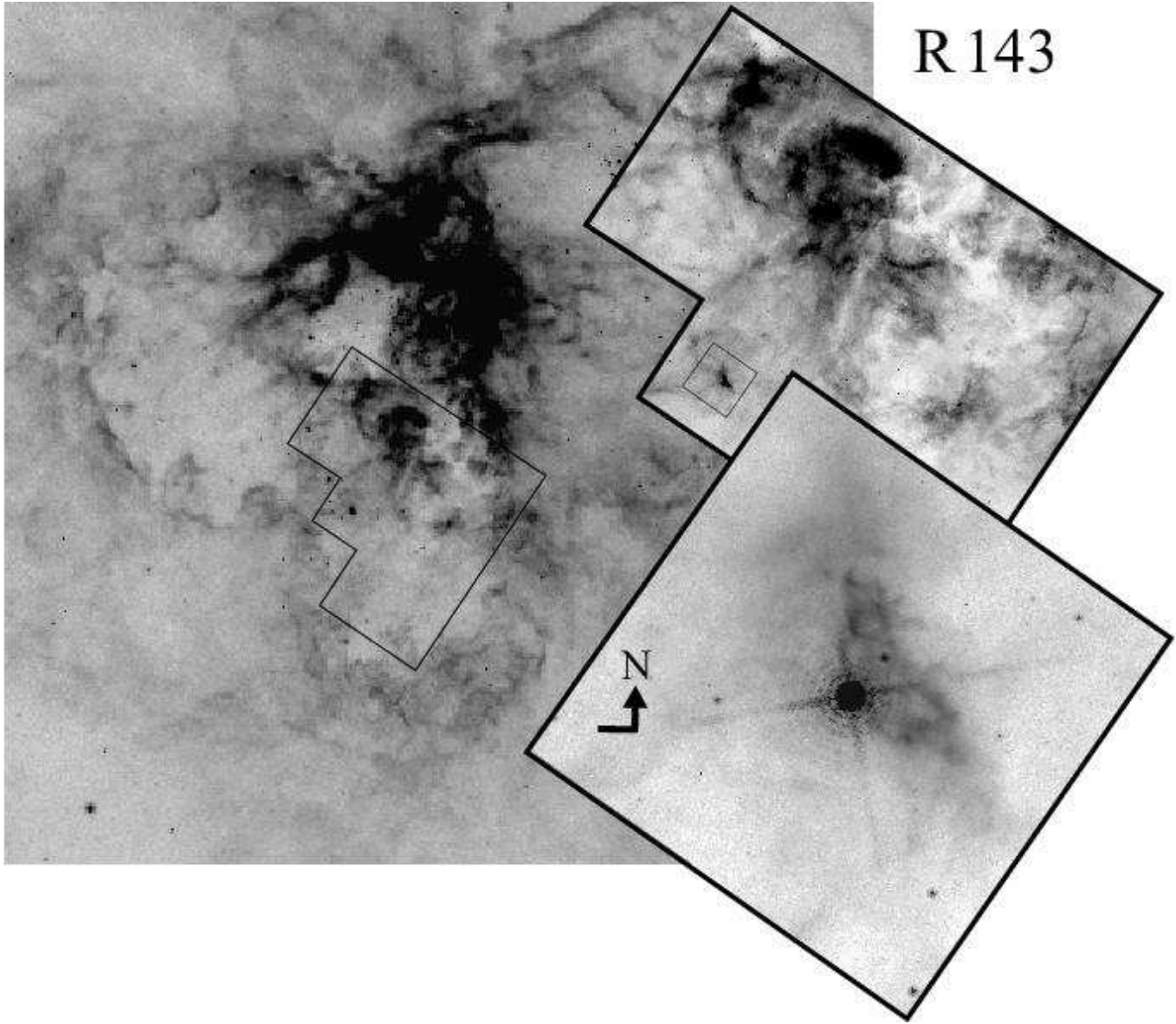}} 
\caption{This collection of images shows the 30\,Dor region
(10\arcmin\,$\times$\,10\arcmin) and in subimages the nebula around R\,143
first in the WFPC2 mosaic and than in the PC of the HST. R\,143 is the bright 
star south-east of the triangular shaped nebula in the small PC image. Only
this triangular shaped nebula is an LBV nebula.
The images illustrated the difficulty to disentangle emission from the 
nebula from that of the 30\,Dor H\,{\sc ii} region.} 
\label{fig:r14330dor}
\end{figure*}

The upper left part of 
Fig. \ref{fig:r14330dor} shows an image taken with the 
0.9\,m-telescope at CTIO (H$_{\alpha}$+[N\,{\sc ii}] filter) 
and, in the upper right, the mosaiked WFPC2 image (F656N), in order
to illustrate the position of R\,143 with
respect to the center of the 30\,Dor complex. The high-resolution 
WFPC2 images underline the difficulty in disentangling 
the LBV nebular emission from the large background H\,{\sc ii} region. 
A section of the PC image as seen in Fig. \ref{fig:r14330dor}
includes the LBV nebula.

The LBV nebula is irregular---triangular 
shaped---and oriented  north to south-west. 
The LBV nebula of R\,143 shows a complex structure consisting of a 
large number of
filaments (lower image in Fig. \ref{fig:r14330dor},
FOV  $\sim$ 10\arcsec\,$\times$\,10\arcsec). 
The LBV nebula is 4\farcs9 in diameter
along the main (long) axis, which corresponds to a linear
size of 1.2\,pc.  The surface brightness is not homogeneous, and 
smaller filaments cross and reach even further
out of the nebula, adding a net-like appearance. The rim of the nebula
is defined through bent or curved structures.  
The nebula is concentrated to the west of the star, and 
no counterpart nebular emission is identified to the east.
The surface brightness of the LBV nebula close to R\,143 was derived to 
$\sim 1 10^{-13}$ergs\,cm$^{-2}$\,s$^{-1}{\rm arcsec}^{-2}$. Feast's
filaments are much fainter with, on average, 
4.3 10$^{-14}$\,ergs\,cm$^{-2}$\,s$^{-1}{\rm arcsec}^{-2}$ .

\subsubsection{The kinematics}

The nebula around R\,143 was observed using three 
different position angles: PA=105\degr, PA=222\degr\ and PA=195\degr.
All slit positions are superimposed on the HST image in 
Fig. \ref{fig:r143slits}.

\begin{figure*}
\begin{center}
{\resizebox{\hsize}{!}{\includegraphics{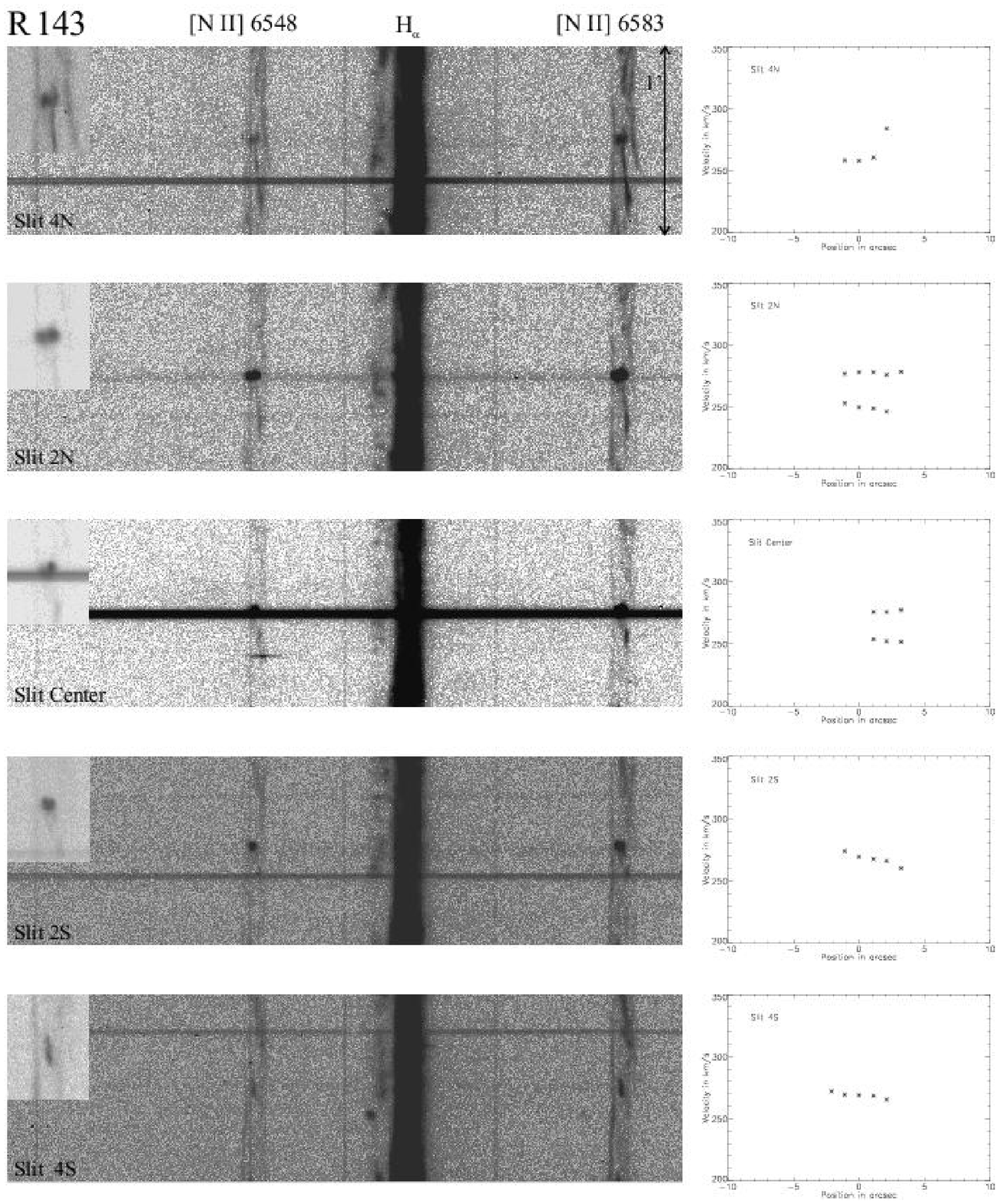}}} 
\end{center}
\caption{Echellograms (left column) and corresponding position-velocity 
diagrams (right column) of our slits for R\,143 (see also
Fig. \ref{fig:echeller143_2}). Measurements are 
with respect to the heliocentric system. North-west (positive positions) 
is up, south-east (negative positions) down. At Slit Center a ghost image
shaped like a dash sign, parallel to the spectral axis, 
appears, superimposed on the 
[N\,{\sc ii}]6548 line.} \label{fig:echeller143} 
\end{figure*}

\begin{figure*}
\begin{center}
{\resizebox{\hsize}{!}{\includegraphics{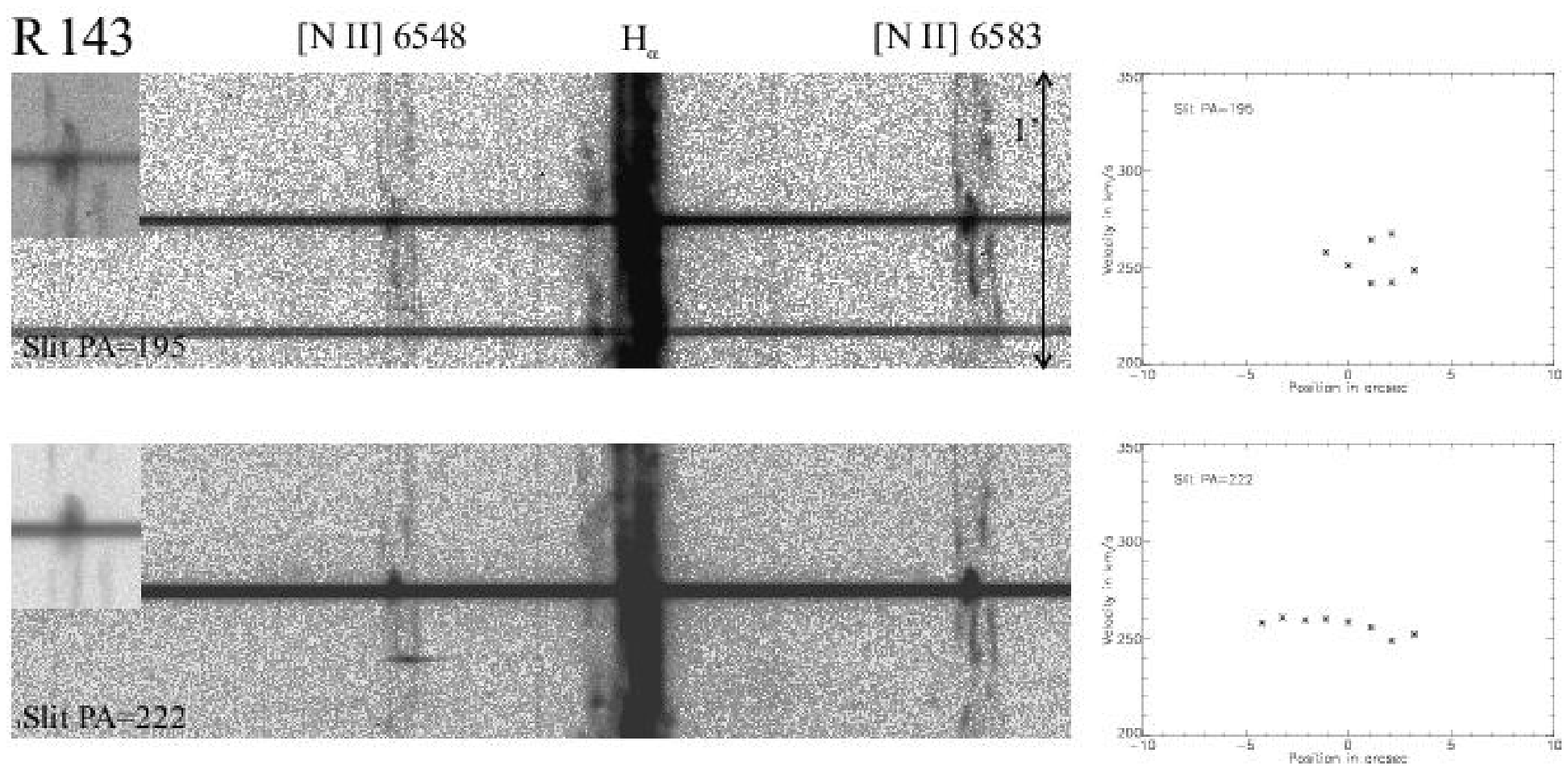}}} 
\end{center}
\caption{Echellograms (left column) and corresponding position-velocity 
diagrams (right column) of our slits for R\,143. Measurements are 
with respect to the 
heliocentric system. South-(west) (positive positions) is up, 
north-(east) (negative positions) down. In both images ghost images 
are on the usual position (see Fig. \ref{fig:echeller143}). } 
\label{fig:echeller143_2} 
\end{figure*}

Using high-resolution long-slit Echelle-spectra, we tried to disentangle 
the emission from the LBV nebula and the background 
30\,Dor H\,{\sc ii} region using kinematic differences and the \NH\ 
ratio---which is expected to be higher in the case of an LBV nebula---as 
indicators. 
The echelleograms  are shown in
Figs. \ref{fig:echeller143} and \ref{fig:echeller143_2}, as well as 
the corresponding $pv$-diagrams.
Slit positions PA=195 and Slit PA=222 were used 
to analyze the kinematics of the
longer arc-like structures found by Feast (1961). 
In Slit 2N, Center, and 2S the inner triangular shaped LBV nebula 
was intercepted. All the echellograms show a large variety of 
structures moving with up to 100\,\kms\ 
velocity differences. This is expected for the region around R\,143 
in the outskirts of the 30\,Dor region, 
which shows a complex expansion structure (Chu \& Kennicutt 1994).
In Slit Center the slit crosses the star R\,143. 
Slightly to the north-east of the stars continuum emission 
(upper direction in the echellogram; Fig. \ref{fig:echeller143}) a  
 [N\,{\sc ii}]-bright knot is visible. The $pv$-diagram 
and the small inset in the echellogram    
show that this knot has two components, one at about 252\,\kms\ and one at
$\sim$ 276\,\kms, a velocity difference of 24\,\kms.
Analogous 
results are found in Slit 2N (where the two velocity components are
even more obvious) and---though not very prominent---in Slit 4N. 
Slit 2S shows only one component declining in velocity from 274\,\kms\ to
260\,\kms. A similar, nearly constant velocity component is
visible in Slit 4S. In all cases, structures with  higher velocities  
are detected, but are  not discussed, since 
they can be attributed to
the emission from the 30\,Dor complex and would only 
lead to confusion if plotted in the $pv$-diagrams. 

\begin{figure}
{\resizebox{\hsize}{!}{\includegraphics{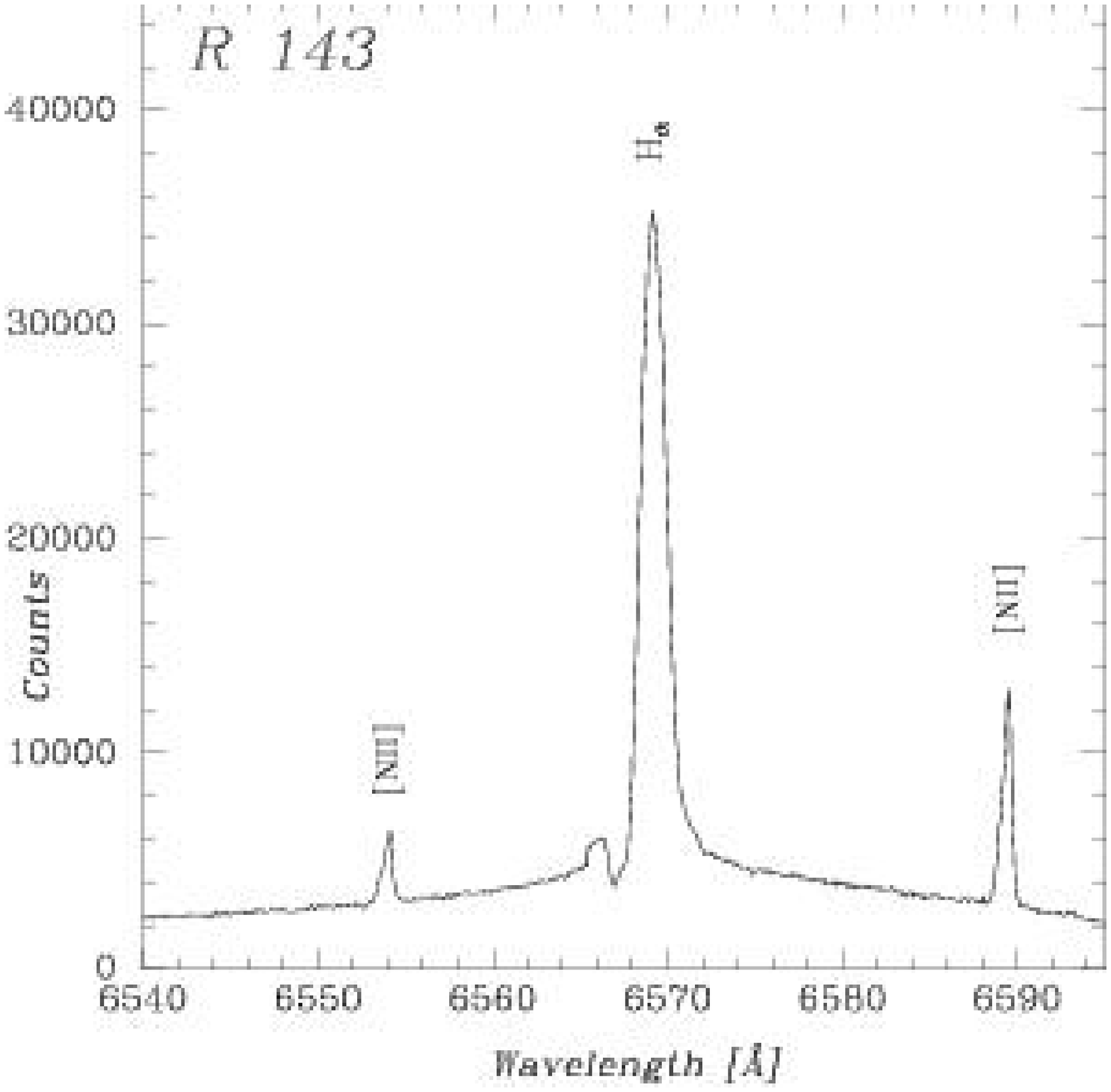}}} \caption{
This plot shows a spectrum of the star R\,143 with a somewhat asymmetric
P Cygni profile
H$_{\alpha}$ line and double peaked [N\,{\sc ii}] lines resulting from the
LBV nebula in line of sight to the star.} \label{fig:r143spectrum} 
\end{figure}

While all the high velocity components in the background emission (not 
included in the 
$pv$-diagrams) and the measurements of the nebula 
in Slit 4S showed a lower ratio of $\sim$ 0.05---normal for H\,{\sc ii}
regions---the measurements in Slit
Center, Slit 2S and 2N reach values of the \NH\ ratio of typically 
0.5 $\pm$ 0.05. 
The maximum we measured is at 0.7 $\pm$ 0.05. 
In Slit 4N the \NH\ ratio still reaches 
up to 0.08. Here due to the seeing, we partly sample the background and
LBV nebula, which results in the slightly enhanced value.
The Slits PA=195 and PA=222 cross the 
star as well as Feasts filaments
(see Fig. \ref{fig:r143slits}). 
The star can be seen in the center in both echellograms (Fig. 
\ref{fig:echeller143_2}) and is surrounded by a small oval structure (the
actual LBV nebula), 
which shows at least in
Slit PA=195 a two velocity component. To the north (downwards in the
echellograms) a longer single velocity structure, slightly bended, appears
(central emission feature).
These emission features (also seen in Slit PA=222) 
are  identified with the filaments
and  show no stronger [N\,{\sc ii}] emission (see also Smith et al. 1998).    

A spectrum extracted from the Echelle spectra of the star itself shows 
(Fig.\ \ref{fig:r143spectrum}) similarities with the spectrum of R\,127.

\subsubsection{Discussion}

Our analysis of the HST images and Echelle spectra support the conclusion that 
Feasts filaments are indeed part of the background H\,{\sc ii} region, and 
that only a roughly triangular shaped high surface brightness nebula 
close to the star manifests parts of an LBV nebula. 
This LBV nebula shows two velocity components separated by 24\,\kms. 
The small size of the nebula makes it hard to identify a clear
expansion pattern in the seeing-limited Echelle spectra. Nevertheless, we
conclude that the nebula moves with two different velocity components and 
shows  a \NH\ ratio higher than that of the
background emission and similar to that of other LBV nebulae. 
The kinematic analysis does not yield other components with  
higher nitrogen emission, which would be part of the LBV nebula.  
Smith et al.\ (1998) find for R\,143 a structure that moves with a 
130\,\kms\ difference from the star, which has, according to them, a 
radial velocity of 348\,\kms. We find that the peak of the stellar 
H$_{\alpha}$ line
has a radial velocity of 295\,\kms\ and is therefore in better agreement with 
measurements of the Si\,{\sc ii} at 4128\,\AA, 4131\,\AA, and Mg\,{\sc ii}
at 4481\,\AA\ by Stahl (private communication), which yield a radial velocity 
for the star of 285.4\,\kms, 285.9\,\kms\ and 287.0\,\kms, respectively.
The large difference of the nebula's and the star's radial velocity of
130\,\kms\ is therefore not supported by our measurements.

The  \NH\ ratio of the LBV nebula we measure is comparable to
the measurements of Smith et al.\ (1998), with the exception that our 
maximum ratio is slightly smaller (0.7) than theirs (0.8). 
The nebula's shape and one-sided location (with respect to the star)
is very unusual and one can only speculate
about various scenarios. With R\,143 situated in the 
30\,Dor region, 
the strong stellar winds, 
the larger density, and the turbulent motion, which is present in such an 
H\,{\sc ii} region, exhibit ideal conditions 
for the easy disruption of circumstellar nebulae.

\subsection{S\,61}\label{section:s61}

\subsubsection{Previous work}

The star S\,61 (or Sk$-67\degr266$) was classified as an O8:{\sc i}afpe extr. 
by Walborn (1977), who noted a similarity of its spectrum  
with that of R\,127. Walborn (1982) puts S\,61 in 
the class of Ofpe/WN9 stars
and detected double peaked nebular lines in the spectrum, indicating that 
S\,61 has an  expanding circumstellar nebula (line split 38\,\kms). He also
noted that nitrogen was $13-16$ times more abundant. Together with the strong
spectral similarity to R\,127 (LBV, see Sect. 
\ref{section:r127}), even though no S Dor-type variability (or eruption)
is known for S\,61, the star is a good LBV candidate. 
Wolf et al.\ (1987) 
found that the wind velocity of S\,61 is 
$v_{\rm wind} \sim 900\,$\kms, and derived a $M_{\rm bol} = -10.3\mag$.    
The first observations to resolve the nebula around S\,61 were published by 
Pasquali et al.\ (1999). The nebula is roughly spherical,
being only slightly 
asymmetric in the center. North of the star (about 1\farcs2)
the nebula is brighter, and the diameter is 7\farcs3 or 1.8\,pc.
Spectra taken of the nebula indicate an expansion velocity, derived from
line-profile fits, of 28\,\kms. 
The nebula and stellar parameters are as follows: 
$n_{\rm e} \sim 400$ cm$^{-3}$, $T_{\rm e} = 6\,120$\,K (Pasquali et 
al.\ 1999), $T_{*} = 36\,100$\,K, 
$R_* = 28\,{\rm R}_{\sun}$, $\dot{M} = 2.2\,10^{-5}\,$M$_{\sun}\,{\rm yr}^{-1}$
(Pasquali et al.\ 1997a), while Crowther \& Smith (1997) 
derive $T_{\rm eff} = 27\,600$\,K, 
$R_* = 33\,{\rm R}_{\sun}$, $\dot{M} = 1.1\,10^{-5}\,$M$_{\sun}\,{\rm
  yr}^{-1}$, $v_{\rm \infty} \sim 250\,$\kms.

\subsubsection{The Morphology from HST images}

\begin{figure}
{\resizebox{\hsize}{!}{\includegraphics{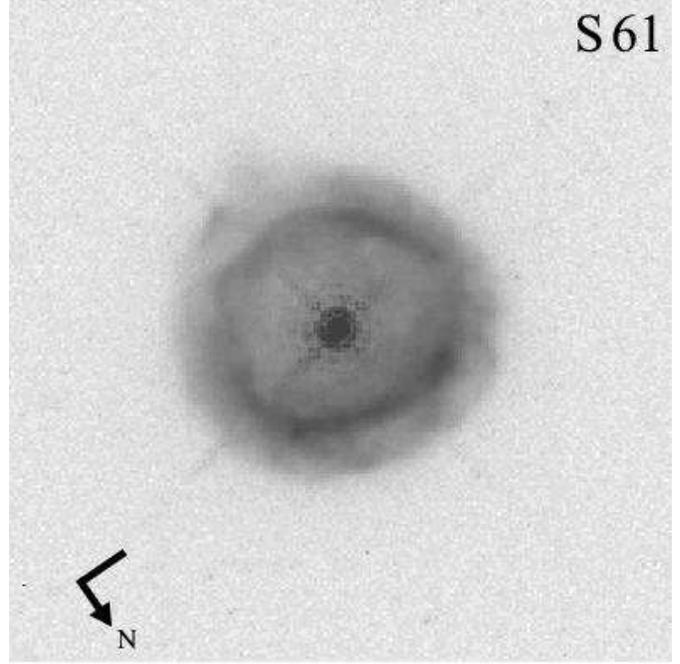}}} \caption{
F656N HST image of S\,61. The field of view
here is about 10\arcsec\,$\times$\,10\arcsec. A north-east 
vector indicates the celestial orientation.
} \label{fig:s61hst} 
\end{figure}

An HST image in the F656N filter is shown in Fig. \ref{fig:s61hst}. 
The FOV here is 10\arcsec\,$\times$\,10\arcsec. 
Previous, ground-based images showed that the nebula around S\,61 is
predominantly spherical. The new HST image supports this morphology 
but reveal more details of the shape of the nebula. 
The nebula consists of an inner brighter ring-like
structure and a larger diffuse emission surpassing this ring in all directions.
The diameter of the inner ring was measured to be 3\farcs4 
(0.82\,pc) at the smallest cross 
section and 3\farcs8 (0.89\,pc) at the largest . The inner ring 
is not perfectly round, but small dips can be seen. The surface brightness is
not homogeneous and is brightest in a section of the ring to the north.
In the ring  the surface brightness varies around 
3\,10$^{-13}$ergs\,cm$^{-2}$\,s$^{-1}{\rm arcsec}^{-2}$, while 
the center of the nebula is fainter with about 
1\,$10^{-13}$ergs\,cm$^{-2}$\,s$^{-1}{\rm arcsec}^{-2}$.
To the south and south-west  the  two largest distortions 
of the nebula are visible within the ring. 
Diffuse emission which shows no clear boundary exceeds the inner 
ring by about 1\arcsec\ or 0.25\,pc. Within this more
diffuse emission smaller filaments are visible.  
These filaments point roughly in a radial direction away from the star.
The emission surpassing the ring is slightly lower in surface brightness than 
the nebula's center (about 
7 $10^{-14}$ergs\,cm$^{-2}$\,s$^{-1}{\rm arcsec}^{-2}$).

\begin{figure}
{\resizebox{\hsize}{!}{\includegraphics{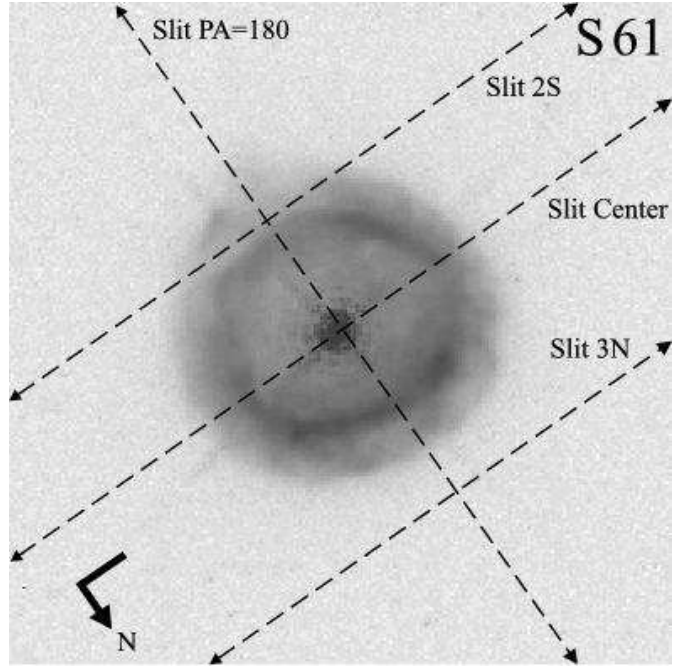}}} 
\caption{ 
Same HST image of S\,61 which now
shows the location of our Slits. 
} \label{fig:s61slits} 
\end{figure}

\subsubsection{The kinematics}

\begin{figure*}
\begin{center}
{\resizebox{\hsize}{!}{\includegraphics{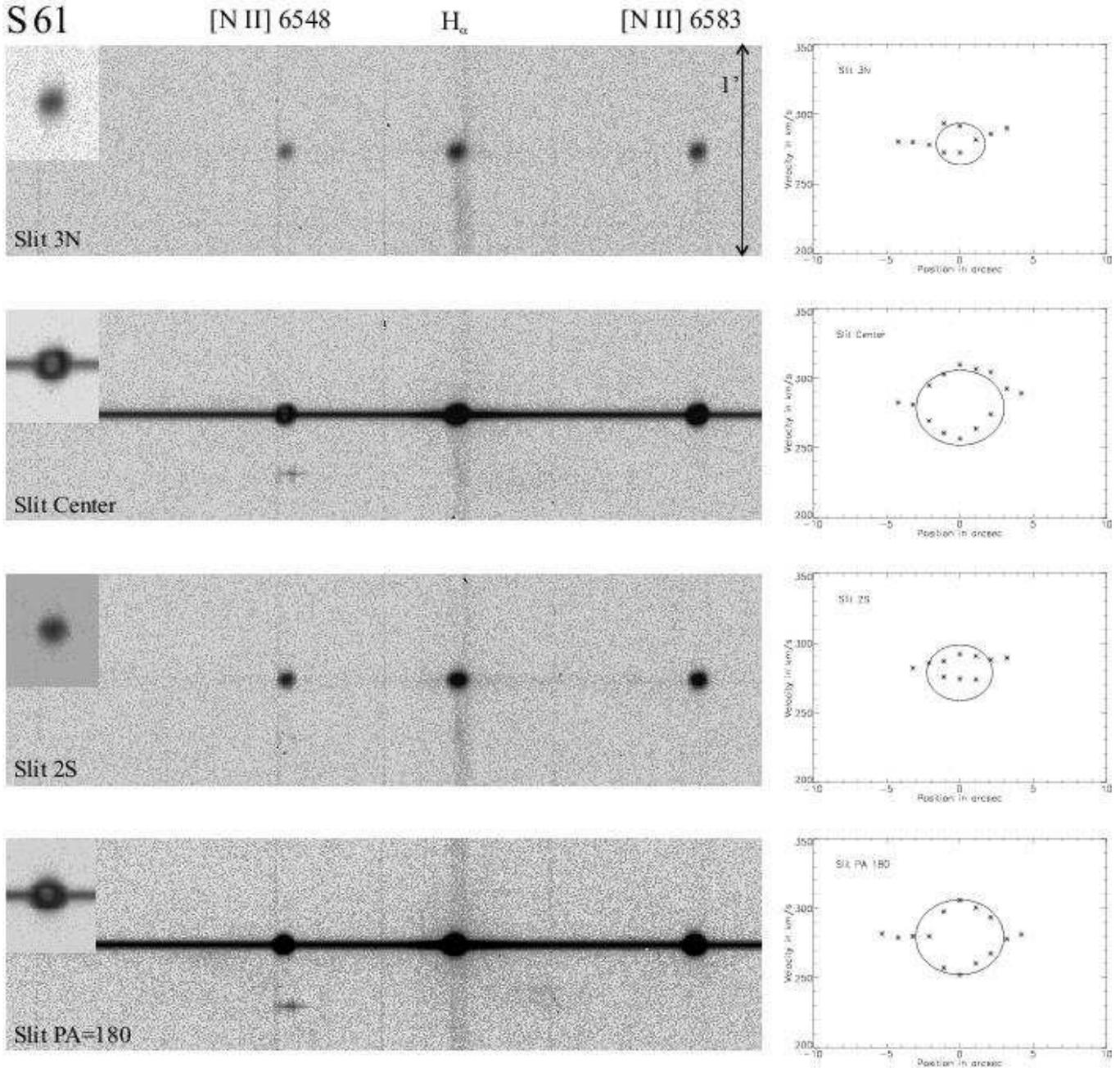}}} 
\end{center}
\caption{Echellograms (left column) and corresponding position-velocity 
diagrams (right column) of our slits. Measurements are with respect to the 
heliocentric system. West (positive positions) is up, east (negative
positions) down. For Slit PA=180 south is up (negativ positions) 
and north down (positive positions).
Slit Center and Slit PA=180 show ghost images. 
All $pv$ diagrams contain the data (asteriks) plus a model of a spherically
expanding shell (line).} \label{fig:echelles61} 
\end{figure*}

Spectra of S\,61 were taken with two different
position angles, perpendicular to each other---a 
PA=90\degr\ (east-west, 3 spectra taken) and PA=180\degr (south-north, one
spectra taken). The positions of the spectra are shown  in Fig. 
\ref{fig:s61slits}, the echellograms and position velocity diagrams in 
Fig. \ref{fig:echelles61}. Even though Slit 3N
is obviously not crossing the nebula, emission is
clearly visible in this spectrum. Most likely the seeing of 2\arcsec\
is responsible for the emission detected at this position. 
In all echellograms 
(see Fig. \ref{fig:echelles61})  either a line split 
is present, or an asymmetrical line shape was found, which indicates a
nearly spherical expansion of the nebula. The maximum expansion velocity 
was measured at position 0\arcsec\ in Slit PA=180 and reaches 26.9\,\kms.
As expected, the expansion is much smaller in the Slits 3N and 
2S---10.5\,\kms\ and 8.9\,\kms, respectively---which occurs because the slit 
intercepts only a section of the sphere. 
Therefore, the diameter of the Doppler ellipse, as well as the expansion 
velocity, is smaller.
The shape of the Doppler ellipse of our data in  Slit Center is 
slightly asymmetric with  
a more redshifted center of expansion at positive offsets (to the west) and 
a more blueshifted center of expansion at negative offsets.
This is illustrated by comparing again the data (asterisks) with the model of 
a purely sphercial expansion (lines) in the $pv$ diagrams.
Besides this small deviation,  the global expansion is in good agreement 
with a spherical expansion. The small deviation in the expansion ellipse might
trace a geometric distortion in the line of slight.

\begin{figure}[h]
{\resizebox{\hsize}{!}{\includegraphics{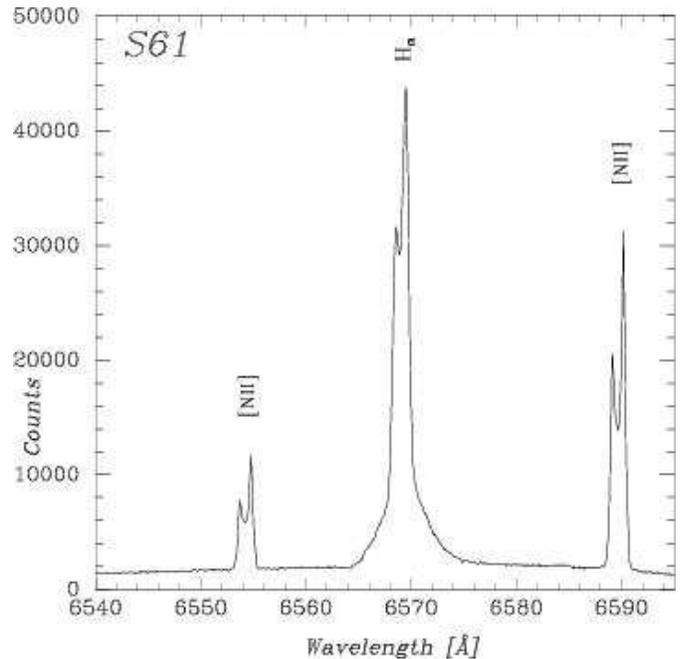}}} \caption{
An extracted stellar spectrum of the star S\,61. The split lines 
visible are emission from the superimposed nebula.
The broad wings seen in H$_{\alpha}$ manifest the broader 
component from the star itself.
} \label{fig:s61spectrum} 
\end{figure}

Fig. \ref{fig:s61spectrum} shows a mixture of an extracted stellar spectrum
plus the superimposed emission from the nebula which lies in the line of
sight to the star. The split nebular lines are very prominent. 
The broader wings in the H$_{\alpha}$ emission most likely show the 
contribution of the stellar line.

\subsubsection{Discussion}

The S\,61 nebula has the best defined spherical structure
among the nebulae discussed so far.
The HST image shows a ring-like 
morphology, disturbed only by small deformations. The same
spherical symmetry dominates the expansion pattern of the nebula.
The nebula around S\,61 nevertheless shows differences from
other classical ring nebulae in being completely surrounded by 
fainter diffuse emission,
which extends beyond the inner ring. 
The seeing in the ground based spectra and the extremely small 
size of the nebula make it hard to 
disentangle the emission from the ring and the fainter outer 
part of the nebula,
since both will be observed and sampled in the spectra simultaneously. 
However, kinematically, the fainter emission can not be very much different
from that of the ring, otherwise an additional fainter component 
should be visible in the spectra with different radial velocity.
From the surface brightness determination 
of the ring and the fainter outer emission in the last section 
we can estimate that the
contribution of the fainter emission  is about 23\%.
Whether this fainter, outer emission is due to an outflow either in all
directions or in the direction that is pointed towards the observer 
(which could form an outflow that expands and is seen projected onto
the inner nebula) is not clear. It might equally be the case that the 
fainter emission was created at the same time as the ring nebula and 
expands uniformly and according to the sphercial shell.
Finally, the outer, faint region of the nebula might be due to an earlier 
high mass loss episode similar to what is found in Planetary Nebulae 
(e.g. Balick \& Frank 2002). 

\subsection{S\,Dor}

\subsubsection{Previous work}

S Doradus (Sk$-69\degr94$, R\,88, HDE 35343) is often considered
the most classical or the prototype of the LBV class.
Its variability has been  known for a long 
time, and early spectra showed variability, too
(Pickering 1897). Numerous studies of its variability have been made, 
for example by Wesselink (1956), Leitherer et al.\ (1985), 
Stahl \& Wolf (1982), Wolf \& Stahl (1990), van Genderen et al.\ (1997b), 
and Wolf \& Kaufer (1997). 
Hubble \& Sandage (1953) reported on the brightest variable stars 
in M31 and M33 and already noted, that S Dor might belong to this group,
later known as the Hubble-Sandage variables. Conti (1984)
suggested that S Dor type stars---the S Dor variables---, the 
Hubble-Sandage variables, and the so called P Cygni
variables could be combined into one group, now the LBVs.
The stellar parameters of S Dor, as measured by Leitherer et al.\ (1985), are
$T_{\rm eff} = 8\,000$\,K, $R_* = 300\,{\rm R}_{\sun}$, 
$\dot{M} = 5\,10^{-5}\,$M$_{\sun}\,{\rm yr}^{-1}$,
and an $M_{\rm bol} \sim -9.7$\mag\ taken from Lamers et al.\ (1998).

Even though S Dor is considered the typical LBV, up to now no 
nebula around the star has been detected. A larger nebulosity 
surrounding the star is most likely part of an H\,{\sc ii} region or 
is a faint superbubble.
S Dor has been in its minimum phase during the last few years; but it recently 
showed an  F type spectrum (Massey 1999, 2000).

\subsubsection{Searching for nebular emission}

\begin{figure*}
\centering
\includegraphics{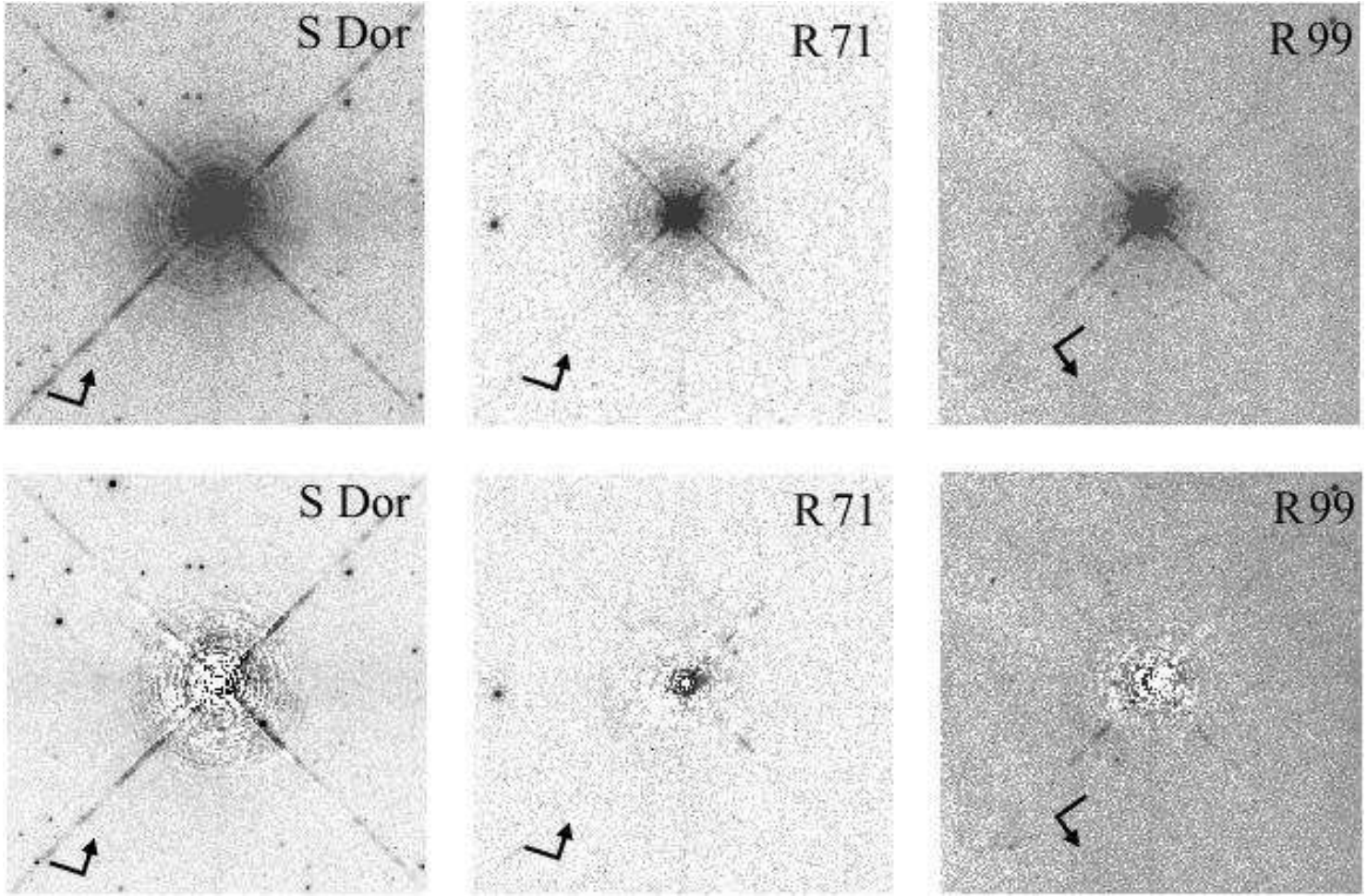} 
\caption{HST 
images of the LBVs S Dor (left panels), R\,71 (middel panels), 
and the LBV candidate R\,99 (right panels). 
The pictures show 15\arcsec\,$\times$\,15\arcsec\ sections of the PC field.
In the bottom panels the 
corresponding images are shown, this time with a 
subtracted PSF.} \label{fig:hstsdorr71r99} 
\end{figure*}

In Fig. \ref{fig:hstsdorr71r99} the upper left panel 
shows an HST image 
(15\arcsec\,$\times$\,15\arcsec) of S Dor in the F656N filter, 
with (top image) and without (bottom panel) the central star. 
In the  bottom left image the stellar PSF was subtracted using 
a PSF generated with Tiny Tim (Krist 1995), to search for 
nebular emission close to the star. 
After the PSF subtraction, no nebular emission around S Dor was found.
The diffuse emission seen
between the two diffraction spikes in the western direction is most likely 
not real and a result of the scattering light, which produces a ghost image at 
that point (see WFPC2  Instrument Handbook, chapter 5.9), that
could not be modeled with Tiny Tim. This residual emission is
visible in all images of Fig. \ref{fig:hstsdorr71r99} at the same position. 

\begin{figure}
{\resizebox{\hsize}{!}{\includegraphics{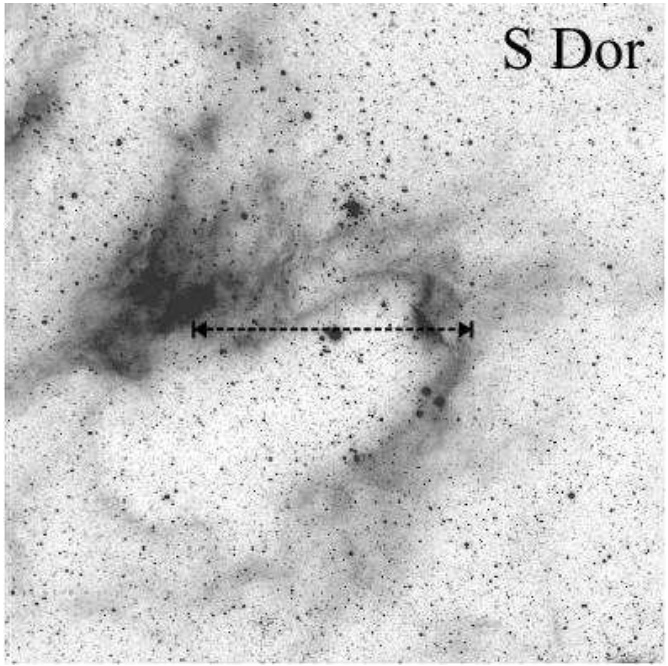}}} \caption{
 H$_{\alpha}$ image taken with the CTIO 0.9\,m-telescope of the larger
 vicinity around S Dor. 
The field of view here is about 10\arcmin\,$\times$\,10\arcmin.
North is up, east to the left. The position of the Echelle slit is indicated
 and drawn to the complete length.
} \label{fig:sdorspalt} 
\end{figure}

An image from the  0.9\,m-telescope (see Fig. \ref{fig:sdorspalt}) was used 
to study the more distant vicinity of S Dor. 
A larger, elliptically shaped, ring-like nebula surrounds S Dor 
with a radius along the  long axis of  5\arcmin\  or about 70\,pc.
The image shows that this structure is clearly defined, and brighter rims
are accompanied by  sections of diffuse emission. To the north-east of
the ring a brighter H$_{\alpha}$ (+[N\,{\sc ii}]) region is visible. 
The emission of the ring around S Dor and the more extended
H$_{\alpha}$ (+[N\,{\sc ii}]) emission  seen in Fig. \ref{fig:sdorspalt} 
are part of the LMC H\,{\sc ii} region N\,119 (or DEM\,L 132). Whether the 
ring-like structure around S Dor was created by S Dor or even resembles 
an LBV nebula is not clear and becomes even more suspicious 
when comparing the
ring with a similar ring structure visible at the north-eastern 
edge of the image (Fig. \ref{fig:sdorspalt}). 

\subsubsection{The kinematics}

\begin{figure}
{\resizebox{\hsize}{!}{\includegraphics{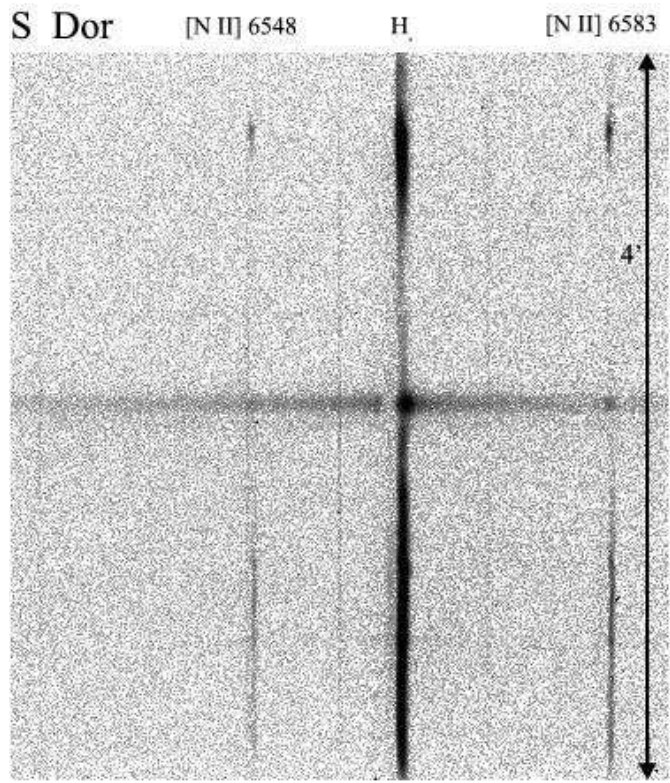}}} \caption{
The  echellogram of the spectrum taken 3\arcsec\ north of S Dor. The full
slit length of 4\arcmin\ is depicted. East is downwards, west up. 
} \label{fig:sdorechelle} 
\end{figure}

\begin{figure}
{\resizebox{\hsize}{!}{\includegraphics{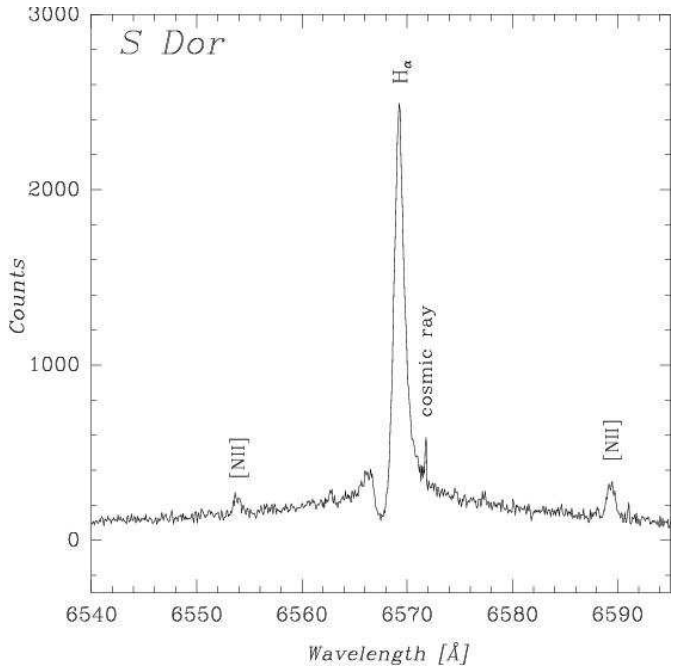}}} 
\caption{Stellar spectrum of S Dor, extracted from the Echelle 
spectrum.  Due to the offset position of the slit, the spectrum 
is of low signal to noise.
} \label{fig:sdorspectrum} 
\end{figure}

An Echelle  spectrum was obtained 3\arcsec\ north of S Dor, running east-west. 
The spatial offset was used to avoid excessive stray light and ghost 
reflections of the grating due to the very bright stellar continnum.
With the bright central star
and a seeing of 2\arcsec, stellar emission is still detected in the spectrum 
(see Fig. \ref{fig:sdorechelle}) and could be extracted (see
Fig. \ref{fig:sdorspectrum}). This somewhat noisy stellar spectrum, 
however, shows the star's H$_{\alpha}$ line with a P Cygni profile 
as well as [N\,{\sc ii}] lines. 
The [N\,{\sc ii}] lines are 
broadended with  an FWHM (corrected for the instrumental FWHM) 
of about 40\,\kms. 
The [N\,{\sc ii}] lines are concentrated on the position of the star 
(see especially Fig. \ref{fig:sdorechelle}) and 
are broader than in  the background [N\,{\sc ii}] lines
(FWHM $\sim$ 19\,\kms), which are visible in other parts of the spectrum. 
Since the 40\,\kms-broadened emission is visible only at the 
position of the star, they indicate nebular emission 
very close to S Dor. A determination of the \NH\ ratio at the star's position 
is not possible since the stellar and nebular H$_{\alpha}$ emission are
superimposed. So far we cannot 
decide whether or not the detected [N\,{\sc ii}] lines indeed manifest a
larger \NH\ ratio, indicative of an LBV nebula.

The echellogram (Fig. \ref{fig:sdorechelle}) is shown in its
full length of 4\arcmin\ to show the H$_{\alpha}$ and [N\,{\sc ii}] 
emission of the surrounding medium.
While we found 
no continuous line split in H$_{\alpha}$,  
which would indicate global expansion, the line from the background is 
also broadened (FWHM $28-33$\,\kms). This value is comparable to the 
turbulent line broadening of an H\,{\sc ii} region (e.g Chu \& Kennicutt 1994). 
Even though no global expansion pattern is visible, 
a second, blueshifted component in the eastern
(lower) part of the spectrum can be identified. While the broad main
component of the gas shows
emission with a radial velocity of 281\,\kms, this second, fainter component
(only detected in H$_{\alpha}$) is moving with 237\,\kms\ and is therefore 
about 40\,\kms\ slower. 
This part of the spectrum coincides with the brightest section of the 
H$_{\alpha}$ emission, visible to the east in Fig. \ref{fig:sdorspalt}.
Most likely several layers of filaments in the H\,{\sc ii} region 
are present here, which move (in part supersonically) with different 
velocities, as e.g. visible in 30\,Dor (see the spectra in the field
of R\,143 in Fig. \ref{fig:echeller143_2}).

Another indication of the origin of the ring around S Dor is 
the \NH\ ratio. As known from other LBVs, this ratio is higher for LBV
nebulae than in normal H\,{\sc ii} regions, due to the CNO processed 
material, assuming that almost no mixing with the
ISM has taken place since the formation of the nebula. 
The \NH\ ratio of the ring as measured in the Echelle spectra is 0.06,
much lower than that of other LBV nebulae and 
comparable to other H\,{\sc ii} regions in the
LMC. If the ring was created by S Dor, it was at least not 
formed during the LBV phase. The \NH\ ratio also makes the
interpretation of the 70\,pc diameter ring as faint superbubble
unlikely, since the typical 
values are between 0.1 and 0.2 (Hunter 1994), due to the diffuse 
radiation field (and/or low velocity shocks) in these objects.

\subsubsection{Discussion}

The HST image of S Dor reveals that, within the limits of 
these observations, no nebular emission is present close to the star. 
The broadened [N\,{\sc ii}] emission found in our spectrum nevertheless 
indicate nebular emission very close to and at the star's position. 
Therefore, it might be that the nebula is less than about 1\arcsec\
or 0.25\,pc in diameter. This conservative limit results
from the PSF subtraction performed on  the HST image. Inspecting  the 
residuals in the PSF subtracted image, we believe that any nebula
larger than that should be visible. The limit was mainly set by the 
bleeding of the central star, which prevents a better subtraction of 
the star's PSF. 

On much larger scales, a ring-like nebula surrounds S Dor
and shows a low \NH\ ratio. This structure was most likely
not formed in the star's LBV phase. With an expected duration of $\sim$ 
25\,000 years of the LBV phase (see e.g., Humphreys \& Davidson 1994), the 
kinematic parameters---the dynamical age is  $\sim$ 10$^6$ 
yrs---support the  arguments that  it is not an LBV nebula.
Nevertheless, we cannot rule out the possibility that the ring around
S Dor was formed by the star's hot wind during the main sequence phase. 
Weaver et al.\ (1977) estimated that  stars 
blow a windblown bubble of comparable size during their main sequence
phase. The ring then is a remnant of such a bubble S Dor blew as a hot 
main sequence star. 
Whether this ring is a main sequence interstellar bubble from S Dor or formed  
purely due to (turbulent) motions within the H\,{\sc ii} region can not be
proven  from our observations. An LBV nebula  can nevertheless be ruled out
due to kinematics---the dynamical age is too large---and the chemical 
composition---no CNO processed material is present.

\subsection{R\,71}
 
\subsubsection{Previous work}
 
Thackeray (1974) suspected that stars classified Aeq and Beq 
(similar as S Dor in its minimum phase)
might indeed be
S Dor type variables, today's  LBVs. The best candidate 
R\,71 (Thackeray 1974; HD 269006, Sk$-71\degr3$)
was classified B2.5{\sc i}ep 
(Feast et al.\ 1960) and 
has a cool expanding shell (Thackeray 1974). 
Wolf et al.\ (1981) derived the following
parameters $T_{\rm eff} = 13\,600$\,K, $R_* = 81\,{\rm R}_{\sun}$, in its
minimum state, and an $M_{\rm bol} \sim -$8.53\mag, rather low for an LBV.
For the cool expanding envelope they find a temperature of about 6\,000\,K.
R\,71 also shows a strong radiation at 10\,$\mu$m, most likely due to a cool 
dust shell (Glass 1984), which is supported by IRAS measurements at 12, 25, and
60\,$\mu$m (Wolf \& Zickgraf 1986). Modeling the dust shell led to 
the following values:  $T_{\rm dust} = 140$\,K, 
$R_{\rm dustshell} = 8000\,{\rm R}_{\sun}$. In the same year, Stahl \& Wolf 
(1986b) detected a broadend, two-component [N\,{\sc ii}] nebular line in the 
spectrum of R\,71, indicating a shell. The de-convolution of the line profile
into two gaussian, yielded a line split of 39\,\kms, which would correspond to 
an expansion velocity of about $v_{\rm exp} \sim \,$20\,\kms.
Using ESO 3.6\,m CASPEC and IUE low resolution spectra, Lennon et al.\ (1994) 
re-determined the stellar parameters of R\,71 after re-calibrating 
the extinction curve for LMC mid-B supergiants. Their values for R\,71 
are:  $T_{\rm eff} = 17\,250$\,K, $R_* = 95\,{\rm R}_{\sun}$, and 
a much higher, for LBVs more typical  $M_{\rm bol} \sim -$9.9\mag.
Only recently  was it found that the dust shell of R\,71 contains 
amorphous as well as crystalline silicate and has a total dust mass of  
0.02 M$_{\sun}$ (Voors et al.\ 1999). 
Even though R\,71 shows the typical light variations 
for an LBV type star (see van Genderen 1979; van Genderen et al.\ 1985, 
1988; Lamers et al.\ 1998)---nebular lines and a dust shell---no optical 
nebula has been detected so far.

\subsubsection{Search for nebular emission}

An HST image of R\,71 in the F656N filter is shown in the   
upper middle panel 
of Fig. \ref{fig:hstsdorr71r99}, as well as the same HST image with the PSF
subtracted (bottom panel). 
After the PSF subtraction, no clear indications of an LBV
nebula were found. Since bleeding affected the images at the star's central
position, and the PSF model does not include bleeding, residual emission of
the star is still visible. Testing our PSF subtraction using differently 
scaled intensities and radii for the PSF models, it was
concluded that a small, point-like emission to the north-west at a distance of 
0\farcs6 from the star might be real but needs to be confirmed.
Note that this emission is not the same
as the false-emission from scattered light reported above, which is more 
arc-shaped and further away from the star. 

\subsubsection{A note on the Echelle spectra}

A spectrum (oriented east-west) of the vicinity of R\,71 was 
taken with an offset of 3\arcsec\ to the north of the star to search 
for nebular emission and expansion. The spatial offset was done for
the same reasons as discussed for the case of S\,Dor.
The spectrum showed no nebular emission at all and is
therefore not shown here, and does not need to be discussed any further. 
Due to the offset, we definitely missed the possible emission knot
detected on the HST image.

\subsubsection{Discussions}

From the measurements on the HST images, as well as our non-detection of
emission in the spectrum, it can be concluded that if R\,71 has 
an LBV nebula, it is either extremely faint and below the detection limit of
the HST image  and in the spectrum, or very small.
If its radius is larger than 0\farcs5, it should be 
resolved and visible in the
HST images where the PSF was subtracted. 
This limit results again from the reliability of the subtracted PSF. At
the LMC distance such a nebula would be smaller than 0.1\,pc. 
A dust shell around R\,71 was observed in 
the IR with a radius of roughly 8000 ${\rm R}_{\sun}$, which would correspond
to 0\farcs00074 (0.00018\,pc) and would therefore, if emitting 
in H$_{\alpha}$, not be visible in the HST images. The detection of a
broadened [N\,{\sc ii}] emission (Stahl \& Wolf 1986b) is indicative 
of an optical nebula, which would expand with less that 40\,\kms.  
%for which a detection with direct imaging has not been successful.
A detection of a point like H$_{\alpha}$ emission north-west of 
the star (on the PSF subtracted image) might be real and could be
part of a small nebula around R\,71.     

\subsection{R\,99}

\subsubsection{Previous work}

The star R\,99 (HD 269445, Sk$-68\degr73$) was
classified as an OBf:pe star (Walborn 1977)
with P Cygni profiles in the highest Balmer series members.
The low-excitation emission spectrum is similar to  that of the galactic LBV 
stars P Cygni and AG Carinae. An IUE study of LMC stars by Hutchings 
(1980) shows that R\,99 has  unusual high reddening 
and peculiar emission lines. He derived a temperature of  $T_{\rm eff} = 
35\,000$\,K. Walborn (1982) and later Bohannan \& Walborn (1989) add R\,99 to
the list of Ofpe/WN9 stars, a stellar class which includes LBVs in their
minimum phase (see Stahl et al.\ 1983, 1984). 

Analysis of R\,99 by Walborn
(1982) showed (somewhat doubtfully) a very weak nebular [N\,{\sc ii}] line at 
6583\,\AA\ with a line
splitting of 61\,\kms. He notes that this line might be due to a 
circumstellar nebula, 
but can also be attributed to the surrounding H\,{\sc ii} region in which
R\,99 is embedded. 
Searching for new LBVs, Stahl et al.\ (1984) combined a large
amount of photometric data to determine variability and the spectral energy 
distributions
of LMC stars. Among them, R\,99 was found to show significant color and 
brightness variations. They derive  an 
extremely  high $M_{\rm bol} \sim -$12.2\mag. 
Even though the star shows typical LBV characteristics, the authors note 
that the amplitude of the star's variation is small compared to the 
classical LBVs.  

Stahl \& Wolf (1987) note
that R\,99 has a spectral  energy distribution similar to HD\,37836 and 
speculate that R\,99 might have a disk. Stahl (1987) searched
for a nebula around R\,99 using ground-based direct 
imaging and the subtraction of the PSF, without success.
Crowther \& Smith (1997) again 
emphasized the peculiarity of the R\,99's
spectrum (showing similarity with the peculiar LBV candidate 
HD\,5980 in the SMC) and its  high terminal wind velocity of about
1000\,\kms. Using NLTE atmosphere models and combining NIR, HST, IUE, and
optical data, Pasquali et al.\ (1997a, 1997b)
obtain the following stellar parameters:  $T_{\rm eff} \sim 34\,000$\,K, 
$R_* \sim 40\,{\rm R}_{\sun}$, and $\dot{M} \sim 3\,10^{-5}\,$M$_{\sun}\,{\rm yr}^{-1}$.
R\,99's variability (van Genderen et al.\ 1998) supports  
its LBV membership  even though it is quite peculiar.
They suggest  that the small variations (see Stahl et al.\ 1984) are likely
due to a Very Long Term S Dor (VLT SD) phase, which shows only a
low amplitude. Van Genderen et al.\ (1998) 
suggest a possible second pulsation mode for R\,99.  

Nota et al.\ (1996a) report the presence of several nebular lines, 
superimposed on the stellar spectrum of R\,99, 
such as [O\,{\sc iii}], [N\,{\sc ii}], 
and  [S\,{\sc ii}]. Using the [S\,{\sc ii}] line ratio, which levels at about 
1.5, the corresponding density lies at around 10-100\,cm$^{-3}$ and is
therefore close to/at the low density limit. 
Their measurements of the H$_{\alpha}$ line at different positions across the
slit identify two major velocity components at $\sim$340\,\kms\ and
$\sim$272\,\kms. The latter agrees with the centroid of the star's 
H$_{\alpha}$ line at 271\,\kms.

\subsubsection{Searching for nebular emission}

\begin{figure*}
\sidecaption
\includegraphics{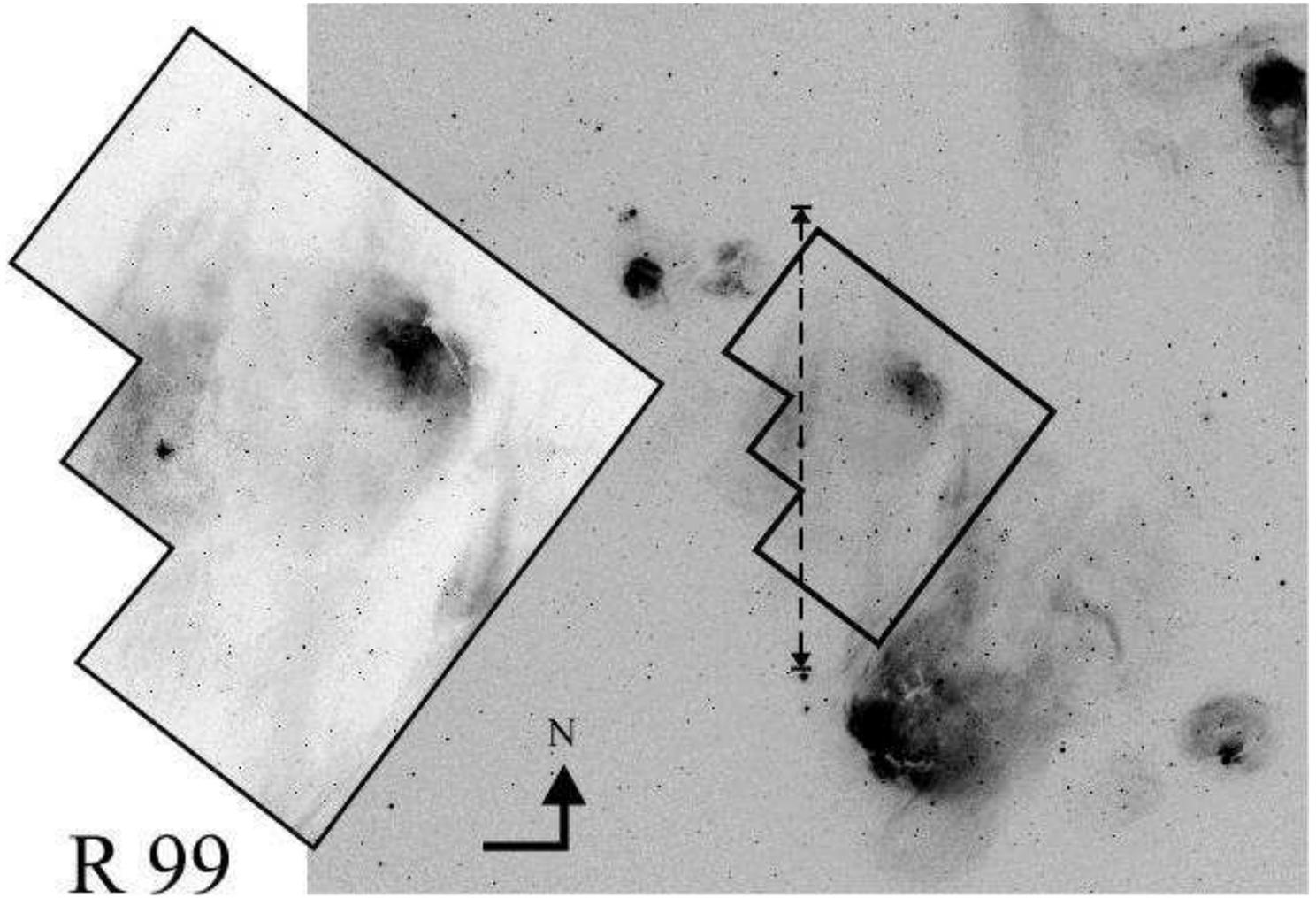}
  \caption{In this figure we combined images taken with NTT-EMMI and the HST
  in H$_{\alpha}$ (+[N\,{\sc ii}] for NTT-EMMI) of the region around R\,99.
The EMMI image is 9\arcmin\,$\times$\,8\arcmin\ large. The footprint of the
WFPC2 shows the enlarged region observed with HST. R\,99 is situated in the PC
section. The full slit length is indicated in the EMMI image. R\,99 
lies close to several H\,{\sc ii} regions in the LMC (DEML\,158, 169, 166a,b)
and is embedded in DEM\,L\,160. Images in the DEM\,L catalog show even 
better the large amount of H$_{\alpha}$ emission present at the star's 
position.
Streaks and filaments in the NTT image hint that cooler gas and dust
are responsible for absorption in these regions. Note that the EMMI
image is displayed at a rather soft contrast compared to e.g. the 
atlas images in Davies et al. (1976). } 
  \label{fig:r99image} 
\end{figure*}

The HST image of the star R\,99 as  in the lower panels in 
Fig. \ref{fig:hstsdorr71r99}  after PSF subtraction does not show any evidence
of emission from a nearby circumstellar nebula.
The image is affected by the same artefacts as those of S Dor and R\,71;
a ghost image shaped like an arc is visible to the north-west. 
In Fig.\ \ref{fig:r99image} an ESO NTT-EMMI H$_{\alpha}$ image is displayed
showing a 9\arcmin\ by 8\arcmin\ region around R\,99. In the same image a
blow up shows the closer area of R\,99 as observed with the HSTs WFPC2 
(long edges about 2.6\arcmin). Both images show that R\,99 is embedded
and surrounded by several H\,{\sc ii} regions which are 
identified with the  help of the plates 
and catalog of Davies et al.\ (1976), see caption in Fig.\ \ref{fig:r99image}.
The images illustrate that most of the 
emission is due to the H\,{\sc ii} regions, and does not show any special 
centering or connection to R\,99 as it would in the case of a 
circumstellar nebula.

\subsubsection{The kinematics}

\begin{figure}
\resizebox{\hsize}{!}{\includegraphics{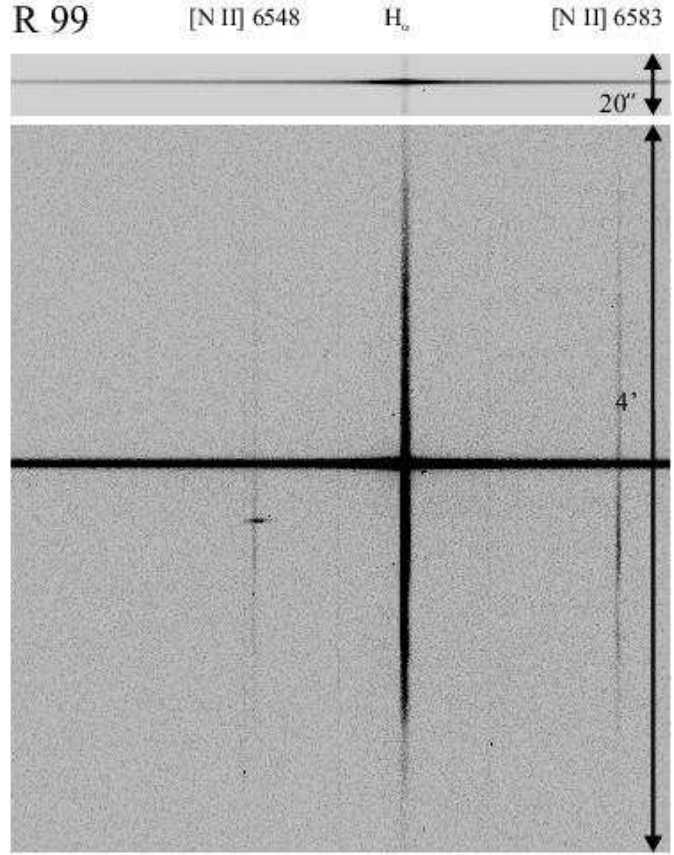}}
 \caption{Echelle spectrum of R\,99 and its surroundings, observed 
with north-south orientation (south is up). Beside the continuum emission
of the stars (see upper panel for a closer view with different intensity
cuts), the spectrum shows H$_{\alpha}$ and [N\,{\sc ii}] emission of the
background H\,{\sc ii} region. Again, a ghost image is present.}  
 \label{fig:r99echelle} 
\end{figure}

\begin{figure}
 \resizebox{\hsize}{!}{\includegraphics{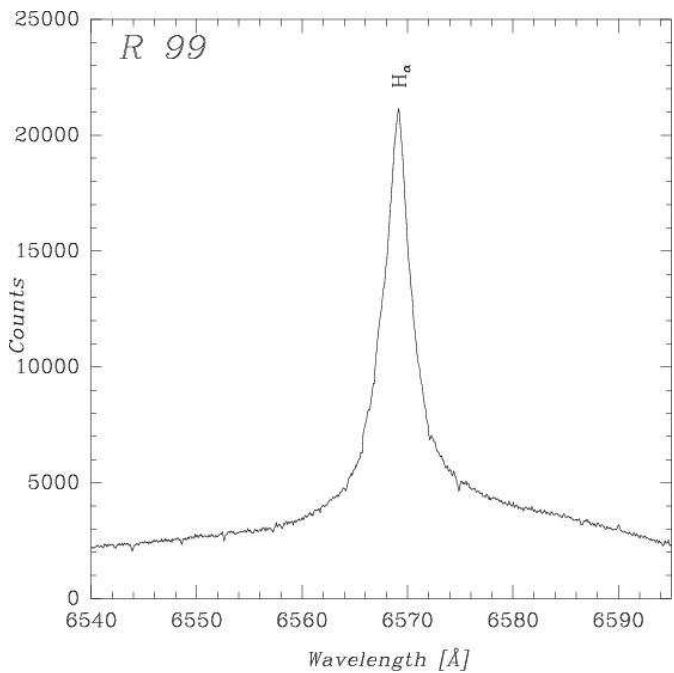}}
 \caption{Extracted stellar spectrum of the star R\,99. The weak line at
   6590\,\AA\ is superimposed [N\,{\sc ii}] background emission (see also 
Fig. \ref{fig:r99echelle}). } 
 \label{fig:r99spectrum} 
\end{figure}

We took one spectrum  on the star, with the slit oriented north-south.
The position and length of the spectrum is also shown in Fig.\ 
\ref{fig:r99image}. The echellogram of the full slit is shown in 
Fig. \ref{fig:r99echelle}, at the top of which only the stellar spectrum is
displayed again with different intensity levels.   
This stellar spectrum was extracted  and can be seen in Fig. 
\ref{fig:r99spectrum}. It 
shows a strong H$_{\alpha}$ emission with two extended and asymmetric wings.
The H$_{\alpha}$ line has a very broad FWHM of 157\,\kms\ and can be best
fitted with a Lorentz profile. The H$_{\alpha}$ emission peak lies at 
286.4\,\kms\ and therefore differs from the measurents of Nota et al.\ (1996)
by 15\,\kms. This difference can be explained by taking the lower spectral
resolution of their  spectra into account. Note, also, that taking a rest
wavelength of H$_{\alpha}$ of 6563\,\AA\ or 6562.8\,\AA\ already makes a
difference of nearly 10\,\kms. Hence, we believe that our measurements 
are not in contradiction but an improvement to the radial velocity of the
star.  A component with 340\,\kms\ as reported by Nota et al.\ (1996)
was not found in our measurements.
Clearly, we detect  H$_{\alpha}$ and [N\,{\sc ii}] emission 
across the entire slit. This is of no surprise since we saw that R\,99 is
embedded in a group of H\,{\sc ii} regions. 
The peak of the H$_{\alpha}$ line of the background lies at
294.5\,\kms\ and is therefore 8.1\,\kms\ more redshifted than the star.
The FWHM of the lines, corrected for the instrumental FWHM, are  28\,\kms\ 
and  18.5\,\kms\ for H$_{\alpha}$ and [N\,{\sc ii}], respectively.
The \NH\ ratio  is about  0.08 $\pm$ 0.02 and typical for H\,{\sc ii} 
regions in the LMC. 
Walborn (1982) speculated whether  the faint  [N\,{\sc ii}] emission 
in his R\,99 spectra was 
from a very faint background H\,{\sc ii} region or a circumstellar
nebula. In the same manner Nota et al.\ (1996) discussed that the emission
lines detected in their spectra can be attributed to a nebula around R\,99.
The Echelle spectra show that [N\,{\sc ii}] is present as part of the
H\,{\sc ii} region. 
The spectrum of R\,99 also shows very faintly (and somewhat doubtful)
the [N\,{\sc ii}] at 6584\,\AA\ (here shifted to 6590\,\AA). 
The velocity of this [N\,{\sc ii}] line is 294.9\,\kms\ and within the errors
identical to the radial velocity of the H$_{\alpha}$ line of the 
background, and rather than the stellar H$_{\alpha}$ line.
We conclude that the [N\,{\sc ii}] line
detected in the stellar spectrum is due to the contamination 
of the spectrum by emission from the background and does not represent a 
nebula around R\,99.
 
\subsubsection{Discussions} 

While on a larger scale (see EMMI images in Fig. \ref{fig:r99image}
and DEM\,L catalog, Davies et al.\ 1976) R\,99 is surrounded by nebula
emission, we find no hints for a circumstellar nebula. The 
PSF subtracted HST images in particular lack evidence of nebular emission
associated with an LBV type nebula of R\,99. 
Earlier detections of nebular [N\,{\sc ii}] emission 
(Walborn 1982; Nota et al.\ 1996) can be accounted 
by the underlying 
H\,{\sc ii} regions---mainly DEM\,L 160---especially in view of the
different centroid radial velocities of the nebular lines with the
stellar H$_{\alpha}$
line. We therefore conclude that the nebular emission detected in spectra 
of R\,99 results from background emission which was acquired at the same time 
of these observations. The densities derived by Nota et al.\ (1996) of
$10-100$\,cm$^{-3}$ are more typical values for H\,{\sc ii} regions
and  support the suggestion, that the nebular emission is part of the 
background H\,{\sc ii}
rather than a circumstellar LBV nebula. If that is the case, the density would 
generally have higher values (several 10$^{2-4}$\, cm$^{-3}$).
The star, however, has an unusally shaped H$_{\alpha}$ line 
(asymmetric Lorentz profile) which is quite broad, again supporting its
exceptional state in our sample.

\subsection{R\,84}

\subsubsection{Previous work}
 
R\,84, also known as HD\,269227 or Sk$-69\degr$\,79, was classified as
spectral type O{\sc i}afpe by Walborn (1977), as he noticed 
``sharp high-excitation emission'', P Cygni profiles, 
and strong hydrogen emission.
Cowley \& Hutchings (1978) and Hutchings (1980) noted that the spectrum of 
R\,84 shows indications for a cool supergiant and concluded that the
spectrum is composite with a  B0e and an M2 supergiant, showing
clear TiO-bands.
The conclusion that the spectrum of R\,84 is a combination of a 
luminous late type and a hot early type star led to the idea that 
R\,84 has a companion star.
Later R\,84 was included in the Ofpe/WN9 sample (see Walborn 1982; Bohannan 
\& Walborn 1989). R\,84 showed strong [N\,{\sc ii}] nebular lines 
which according to Walborn (1982) are ``quite strong...and although there is 
no clear velocity structure, they are broader than single, unresolved nebular 
lines would be''.

Glass (1984) reports a strong IR excess but finds no TiO-bands; he 
identifies R\,84 as a composite Wolf-Rayet star plus late type 
supergiant. Studies by Stahl et al.\ (1984) confirm the large IR excess,
but note that the (K-L) value is too high to 
result only from a late-type companion.
They propose that a circumstellar dust shell surrounds R\,84 and 
strengthen this argument with the peculiar variation of the shape of the
continuum of the star and a variable 2200\,\AA-feature, which could be 
due to dust formation. Stahl et al.\ (1984) derived  the following 
stellar parameters for R\,84: $T_{\rm eff} = 25\,000$\,K, 
$R_* = 33\,{\rm R}_{\sun}$, and $M_{\rm bol} \sim -$9.3\mag. 
R\,84 is accordingly described as an S Dor variable with  small 
amplitude  variations.

Further evidence for a circumstellar shell around R\,84 came from a 
broad, most likely non-interstellar Na{\sc i} D line and 
the resolved (FWHM = 37\,\kms)  [N\,{\sc ii}] lines (Stahl \& Wolf 1986b). 
Direct imaging observations to resolve a circumstellar nebula were 
not successful (see Stahl 1987). Wolf et al.\ (1987) raised
doubts about the fact that the companion star is responsible 
for the star's peculiarities, 
showing that the UV spectra of R\,84 is that of an O9.5{\sc i}ab star 
and quite similar to that of S\,61  (see 
Sect. \ref{section:s61}). Several atmospheric analyses of R\,84 (Schmutz et 
al. 1991; Crowther et al.\ 1995; Nota et al.\ 1996a; Pasquali et al.\ 1997a)
in the few last years derived consistent results for the stellar parameters, 
also comparable to those mentioned above, but favored
different evolutionary scenarios for R\,84. Schmutz et al.\ (1991) proposed
that R\,84 had gone through a red supergiant phase, Crowther et al.\ (1995)
compared it with the dormant LBV R\,71, and Pasquali et al.\ (1997a) 
identified R\,84 as a quiescent LBV.  

A recent analysis with ESO NTT SUSI images and 
using ADONIS adaptive optics by Heydari-Malayeri et al.\ (1997) 
showed that, if a
companion star is in the line of sight to R\,84 (or a close binary),
it must be closer than 0\farcs12. Van Genderen \&
Sterken (1999) concluded that R\,84 shows no signs of variability 
and earlier reports of variations might be due to 
faint field stars in the photometric 
aperture. They classify R\,84 as an ex-/dor\-mant LBV.
Obviously, the question of R\,84's evolutionary state and history
is not settled yet.
We include R\,84 in our set of LBV candidates, and discuss the possibility 
that R\,84 is an LBV and whether it is surrounded by a circumstellar shell. 
No high-resolution or HST images of R\,84 are available at this point.

\subsubsection{Images}

\begin{figure*}
\sidecaption
\includegraphics{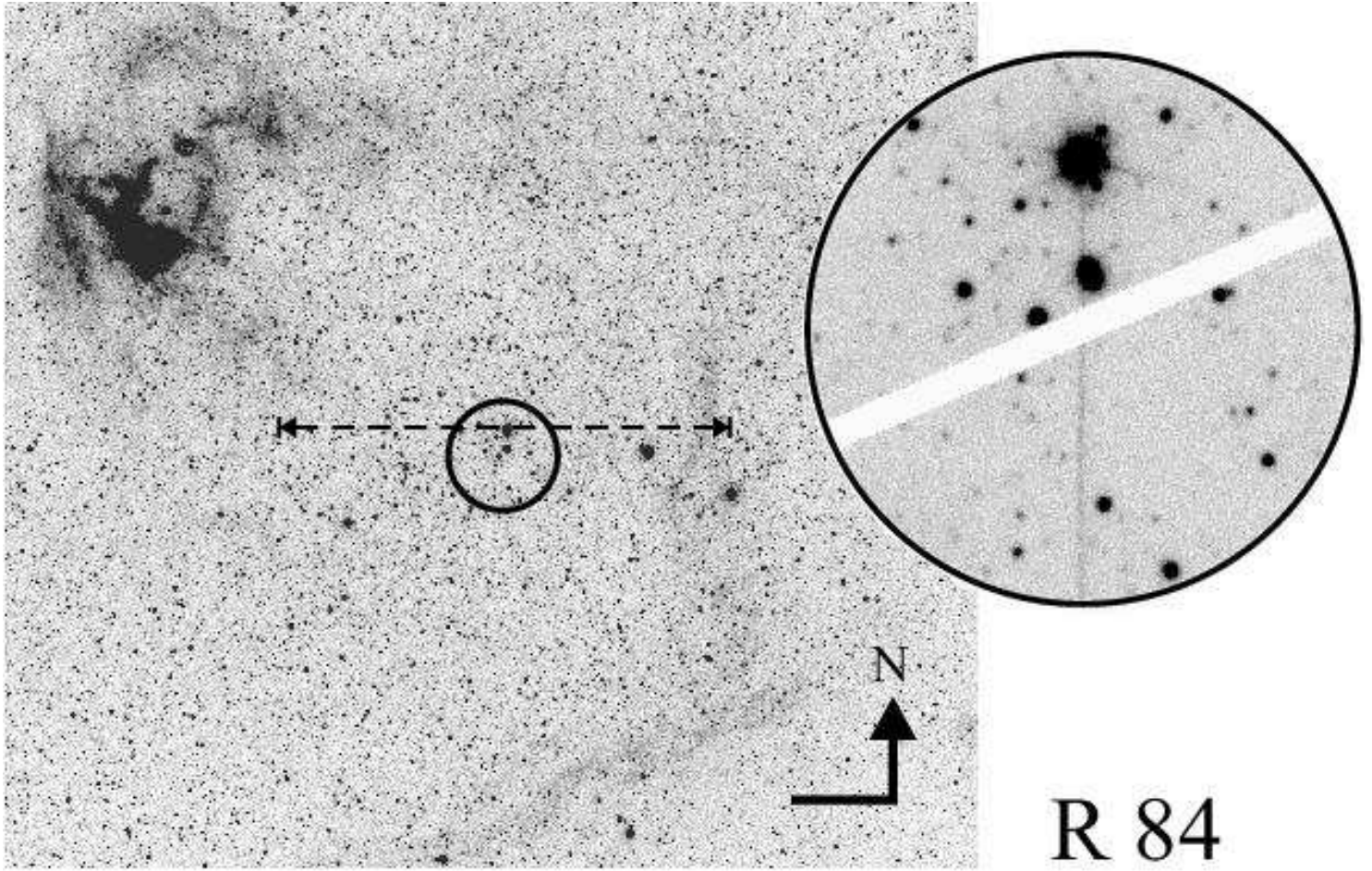} 
\caption{This figure shows a composite H$_{\alpha}$ image constructed 
from ESO NTT-EMMI  and higher resolution ESO NTT-SUSI observations 
of the area around  R\,84. The EMMI image (left) is 
9\arcmin\,$\times$\,8\arcmin, the inlet with the SUSI image is
somewhat larger than 50\arcsec. 
The SUSI image was taken using a coronographic mask, which blocks part of the
emission in the field. R\,84 is the bright star in the northern part of the
SUSI image. Due to its brightness, charge transfer effects are seen starting
at the star's position and extending down the chip as a long streak.
In addition, the
slit position with the full 4\arcmin\ length is indicated in the EMMI image. 
The larger H\,{\sc ii} region north-east of R\,84 is part of DEM\,L\,113.
Faint filamentary nebular emission (DEM\,L\,110) is also seen west 
of R\,84, stretching roughly north-south. 
The high density of stars visible particularly in the EMMI image is due
to the location of the field near to the stellar bar of the LMC.} 
\label{fig:r84image} 
\end{figure*}

Since no HST images are available, we used ESO NTT-SUSI archive images with a 
seeing of 0\farcs75 and inspected these images for possible nebular emission
associated with the star. A NTT-SUSI image of R\,84 is shown in Fig.\
\ref{fig:r84image}, together with an image on a larger scale
(9\,\arcmin\,$\times$\,8\arcmin) taken with the NTT-EMMI. 
In both images an H$_{\alpha}$ (plus [N\,{\sc ii}])
filter was used. The EMMI image 
(Fig. \ref{fig:r84image}) shows that R\,84 is in the vicinity of the 
H\,{\sc ii} region DEM\,L 113 (Davies et al.\ 1976), which 
is partially seen in the upper left corner of the EMMI image. 
To the west (right) of R\,84 very faint H$_{\alpha}$ filaments 
are visible, which are part of DEM\,L 110 and are roughly oriented
north-south before they kink to the east. Comparing the EMMI images with the 
images of the DEM\,L catalog shows that very faint emission also lies 
between the the filaments (DEM\,L 110) and DEM\,L 113 and would hence be 
present at the position of R\,84.
In the ESO SUSI image, which shows an area of about 13\,$\times$\,13\,pc, 
no obvious H$_{\alpha}$ emission was found. R\,84, which is the brightest star
in the northern part of the image, nevertheless show no perfectly round
structure. Some streaks point away from the star, most of them due to 
diffraction spikes and charge transfer problems (long line streching down).

\subsubsection{The kinematics} 

\begin{figure}
\resizebox{\hsize}{!}{\includegraphics{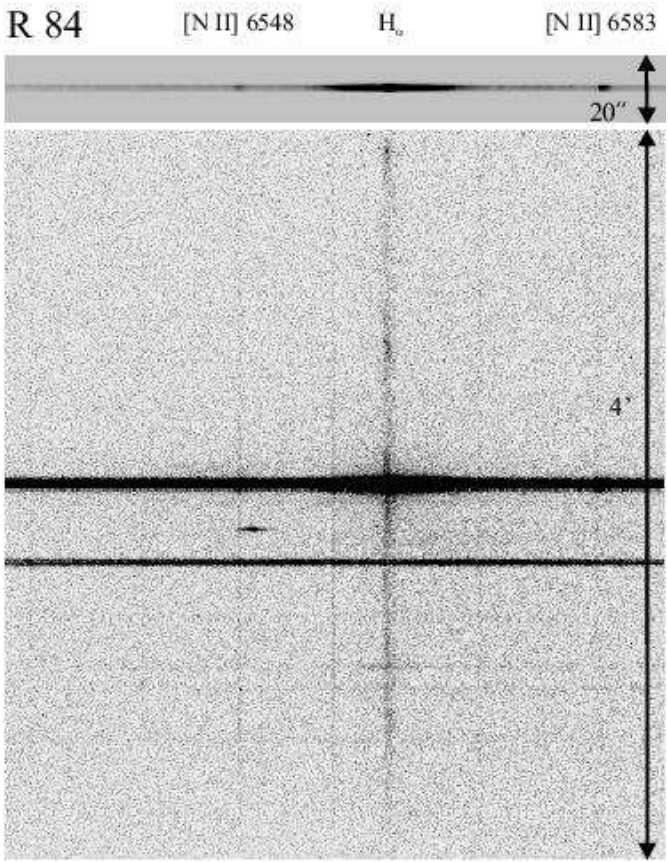}}
 \caption{The  Echelle spectrum of the star R\,84 and its vicinity, slit
   orientation is east-west. West is up. The upper panel illustrates
   the stellar continuum spectra in which the broad H$_{\alpha}$ and [N\,{\sc
   ii}] lines show up at higher intensity. Across the slit some
 background H$_{\alpha}$ emission was found, but the background 
[N\,{\sc ii}] is barely visible.} 
 \label{fig:r84echelle} 
\end{figure}

\begin{figure}
\resizebox{\hsize}{!}{\includegraphics{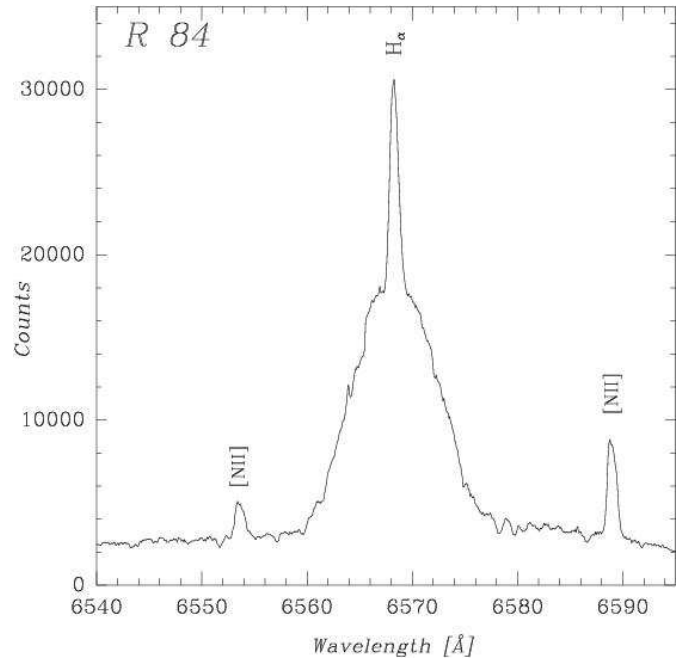}} \caption{
An extracted spectrum of the star R\,84. The nebula [N\,{\sc ii}] lines 
are doubled as is the narrow H$_{\alpha}$ line.
} \label{fig:r84spectrum} 
\end{figure}

We took one Echelle spectrum centered on the 
star (see Fig.~\ref{fig:r84echelle}) 
and one offset by 3\arcsec\ north, both with an east-west orientation. 
Since the
offset spectrum does not contain any additional information, it is not 
shown and will not be discussed here.
Of the LBV candidates in the LMC,  R\,84 is the strangest object.
In its  spectrum the H$_{\alpha}$ line  shows a broad and a narrow 
component 
(see extracted spectrum in Fig. \ref{fig:r84spectrum}). 
The broad component has an FWHM of 435\,\kms. The narrow component
is split, as are the [N\,{\sc ii}] lines. The  peaks of the split
have  radial velocities of 241\,\kms\ and 
265\,\kms, and  are therefore separated by 24\,\kms\
(in agreement with measurements by 
Stahl \& Wolf 1986a). 
The narrow split H$_{\alpha}$ line seems superimposed on the 
broad  component. The narrow and broad component can be fitted (and
subtracted) with just two gaussians. 
The split [N\,{\sc ii}] emission is quite bright, and  
if we compare it with the narrow H$_{\alpha}$ line, we obtain an
\NH\ ratio of about 0.5 $\pm$ 0.1.
It is probable that different 
origins are responsible for the broad and the narrow components.
The narrow component  most likely results from
nebular emission, while the broad
component represents H$_{\alpha}$ emission from the star.
The echellogram (see Fig. \ref{fig:r84echelle}) shows that the 
nitrogen emission is detected only at the position of the star (compare full
spectrum with insert to the top, which contains only the star, but with
different brightness cuts). While H$_{\alpha}$ emission of a faint
background H\,{\sc ii} region is seen across the entire slit, 
[N\,{\sc ii}] emission is not, or is only barely visible. It is at least 
much fainter than the  [N\,{\sc ii}] emission seen superimposed on the star's
continuum. Summing up all the  [N\,{\sc ii}] emission for a better signal to
noise, we obtain an \NH\ ratio of 0.1 for the background, much less than the
ratio at the star's position.  
In the echellogram we can see that the 
H$_{\alpha}$ line 
is slightly brighter closer to the star
(up and down in the echellogram, Fig. \ref{fig:r84echelle}). 
This brighter region has a diameter of
23\arcsec. At the same time, no bright
[N\,{\sc ii}] is seen, and the \NH\ ratio stays the same. 

\subsubsection{Discussions}

We find that bright [N\,{\sc ii}] emission is concentrated 
at the position of the star and supports the presence of a 
very small circumstellar nebula around R\,84 which is not
spatially resolved. 
The size of such a nebula can only be estimated to be smaller than 
2\arcsec\ (about 0.3\,pc). This limit is given by the
seeing we had in the Echelle observations. 
The NTT-SUSI image shows some faint protrusions which could be part of a
small circumstellar nebula, but the quality of the data precludes a 
definite statement on details of the nebula structure. 
The spectrum indicates that the expansion velocity of this nebula
is about 12\,\kms, as derived from the separation of the narrow lines.
The \NH\ ratio is in good agreement with values found for LBV
nebulae. 
Faint nebular emission  was also observed above or below the star in the
spectrum but shows a much lower \NH\ ratio, typical 
for an H\,{\sc ii} region. 
The broad component in the spectrum of R\,84---only visible in 
H$_{\alpha}$---has not been seen in other LBVs and opens questions about 
whether it is a new spectral feature unique to R\,84 or could possibly be
related to a binary. The narrow  H$_{\alpha}$ component is most 
likely from
the nebula that also forms the [N\,{\sc ii}] lines.
The broad H$_{\alpha}$ component could be the stellar emission
broadened due to a very strong stellar wind.

\section{Summary of previous results for other LBV candidates in the LMC}

Observations and analysis for the two LBV candidates 
S\,119  and \sk\ have been published earlier
in  Weis et al.\ (2003) and Weis \& Duschl (2002), respectively.
In the following we will briefly summarize these results to put these objects 
in context with the results of this paper.

Note in this context that  we use the term outflow from a
nebula in the sense of outflowing material from a disrupted shell.
Outflow here is used to describe the gas which flows out of the nebula,
it is {\it not} used for the stellar wind outflowing from the star.

\subsection{S\,119}

S\,119 (Sk$-69\degr$\,17, HDE\,269687) is classified as an Ofpe/WN9 
star (Bohannan \& Walborn 1989), and the latest determination of the stellar
parameters yield $T_{\rm eff}=26\,200/27\,000$\,K, 
log\,$L = 5.76/5.80$\,L$_{\sun}$, and 
log\,${\dot M} =-4.87/-4.92$\,M$_{\sun}$yr$^{-1}$ (Crowther \& Smith 1997).
Different values are obtained for two different models to take the nebula 
contamination into account.  The terminal wind velocity is 400\,\kms (Pasquali
et al.\ 1997a). 
Nota et al.\ (1994) were the first to resolve a  7\arcsec\,$\times$\,9\arcsec\ 
large nebula (1.9\,pc $\times$ 2.1\,pc) associated with 
S\,119. We analysed this nebula (Weis et al.\ 2003)
using HST imaging and 4\,m long-slit
high-resolution spectra,  in the same set-up as descibed in this work.
The images show that the nebula is nearly spherical with a diameter of 
7\farcs5 corresponding to 1.8\,pc. The shell is brightest in the east. 
From the south-east toward the north-west part of the nebula several filaments 
are stretching out, the longest extending 1\farcs81 beyond the nebula's shell. 
Our long slit observations show an expansion of the nebula main shell with
a maximum expansion velocity of 25.5\,\kms. Beside the 
spherical expansion of the central part of the nebula, we showed 
that---as the filaments in the images indicate---material is streaming out 
with much higher radial velocity. The  
highest velocity detected in the outstreaming gas 
is with 283\,\kms\ about 130\,\kms\ faster than the center of
expansion (with 156\,\kms). 
The nebula around S\,119 is not completely closed and shows outflow.
Such an outflow might results, e.g., from Rayleigh-Taylor instabilities in the
nebula, density gradients in the ambient medium or the onset of an 
asymmetric faster stellar wind. 
Finally, note that the location of
S\,119 in the disk of the LMC is still under 
debate since the star's radial velocity 
is much slower than expected (Nota et al. 1994; Danforth \& Chu 2001; 
Weis et al.\ 2003). Hence, the nebula's size might be different, 
giving a possibly shorter distance.

\subsection{\sk}

\sk\ was recognized as an H$_{\alpha}$ emission-line star by Bohannan \& Epps 
(1974) and later classified as O9f by Conti et al. (1986).
Thompson et al.\ (1982) derived
an effective temperature $T_{\rm eff} = 30\,300\,$K, an absolute
bolometric magnitude $M_{\rm bol}=-9\fm72$, and a visual one 
$M_{\rm V}=-6\fm77$. We found (Weis et al.\ 1995)
that the star \sk\ is surrounded by a nebula with a diameter of 
18\arcsec\ or 4.5\,pc. The nebula expands with roughly 14\,\kms.
Several analysis and studies of this object (Weis et
al.\ 1995, 1997b; Weis \& Duschl 2002) let us suspect that this object is an
LBV candidate due to its brightness, UV spectra (see Smith Neubig \&
Bruhweiler 1999), and the large nebula, which is especially strong 
in the [N\,{\sc ii}] lines. 
The \NH\ ratio of 0.7 is comparable to those of other LBV nebulae
(e.g., AG Car: \NH $\sim$ 0.5, de Freitas Pacheco et al.\ 1992). 
Recently (Weis \& Duschl 2002), we detected that \sk\ also  
exhibits a large scale outflow similar to that seen in S\,119.
This outflow is seen as a  7\farcs1 (1.7\,pc) large filament extending
to the north of the nebula.  
The filament moves about 21\,\kms\ faster than the center of expansion
of the nebula.

\section{Summary of this work}

From all other stars analyzed here,  R\,99 was the object that  
showed the least evidence of a nebula. Neither the PSF subtracted 
HST images, nor the Echelle spectra show any indications of nebular emission. 
If an LBV nebula  surrounds R\,99, it must be at least smaller or much fainter 
than what we can measure, giving our detection limits. 
The LBV/LBV candidates R\,127, R\,143, S\,61, S\,119, and \sk\ 
all show  nebulae which 
are several arcseconds in diameter. We find that R\,127's nebula is 
fairly spherical with two triangular-like attachments, the  
caps. The kinematics of the nebula gives hint for bipolarity.
In R\,143 only a smaller irregularly---nevertheless 
approximately triangularly---shaped nebula with net-like structures
is identified as an LBV nebula. Our spectra
show two  velocity components, separated by 24\,\kms\ associated 
with this LBV nebula.
For S\,61, the nebula's morphology is predominantly spherical ($\sim$3\farcs6),
with fainter emission that---in projection---surpasses
the borders, the global expansion velocity is about 27\,\kms.
Analyzing images of S Dor, R\,71, and  R\,84, we found no clear indication for 
nebular emission. Filamentary, in the case of S Dor even bubble-like emission 
visible in H$_{\alpha}$ found close to S Dor and R\,84 are 
part of the H\,{\sc ii} regions 
or diffuse hot gas in the LMC, but are not relics of 
the stars' LBV phase. However, note that all these objects, superimposed on the
stellar spectrum, show nebula [N\,{\sc ii}] lines which are broadened (S Dor) 
or split (R\,71, R\,84).  These lines are
indicative of nebular emission in connection with these stars. The higher line
ratio and spatial extend  might result from LBV nebulae close to the 
stars which 
we can not resolve spatially.   

\section{General Discussion and Conclusions}

In this final section we will compare the parameters of the LMC LBV nebulae
studied here with those of the known Galactic LBVs. Therefore, we summarize
the morphologic and kinematic parameters of both groups in
Table \ref{table:lbvn}. For a detailed description and sources of the
parameters, we give examples of the most recent literature 
which, however, is far from complete. In Table \ref{table:lbvn} we quote the 
size, expansion velocity, and morphology of the nebulae. In some cases, 
where no clear expansion ellipse could be traced, 
not the expansion velocity but the line split is given, which is marked with
(split). For S Dor the line was not split, so the FWHM is given instead.

\subsection{Morphology and Nebula Sizes}

The different morphologies and sizes of nebulae around LBVs and LBV candidates 
in the LMC are particularly conspicuous. 
The nebulae are spherical, 
bipolar, irregular, or show outflow. Also, they appear to be grouped into 
two different size classes. Either they are quite large (1\,pc or larger) or
they are very small (upper limit from the resolution of our data 0.3\,pc).
Of course, we have to keep in mind that selection  effects play an
important role. 
In the LMC we predominantly detect the largest nebulae  since
those smaller than about 0.5\,pc correspond to an apparent size of 
about 2\arcsec\ in the LMC and are therefore much harder to find in 
surveys, for instance. 
This is  especially the case for stars which are not identified as LBVs
so far because they are in a dormant state. 
The LBV nebulae  in the LMC are in general about the same sizes (perhaps
slightly larger) as those in our Galaxy. From all nebulae which have been
resolved so far this shows that they cover sizes 
between 0.15 (HD 168625) and 4.5\,pc (\sk).
Taking again all resolved nebulae into account, we can estimate an 
average size for LBV nebulae  of $\sim$1.25\,pc.

LBV nebulae in the Galaxy and the LMC come in the same variations of
morphologies. In particular 
bipolarity---at least to some degree---in LBV nebulae
is present in nebulae in both galaxies. Interestingly,  
we find that bipolarity is found in nebulae of very different sizes. 
A bipolar nebula {\it par excellence}  is seen in the 
Homunculus  around $\eta$ Car, but also larger nebulae like those
around HR Car and R\,127 in the LMC have bipolar components. Conclusively, 
bipolarity is neither restricted to size, nor connected to the host galaxy (or 
as a consequence: metalicity). 

\begin{table*}
\caption[]{Comparison of the parameters of Galactic and LMC 
LBV nebulae. Numbers separated by slashes indicate that the 
nebula consists of two 
parts. The sizes are given as diameters or the minimum and maximum 
extension (smallest and largest axes).}\label{table:lbvn}
\begin{center}
\begin{tabular}{ccccccc}
\hline
\hline
LBV & host galaxy & size & $v_{\rm exp}$ & morphology& references \\
& & [pc] & [\kms] & \\
\hline
\hline
$\eta$ Carinae & Milky Way & 0.2/0.67 & 600/$10-2000$ & bipolar & Weis (2001); Weis et al.\ (2001) \\
HR Carinae & Milky Way & 1.3\,$\times$\,0.65 & $75-150$ & bipolar & 
Weis et al. (1997a); Nota et al.\ (1997) \\
P Cygni & Milky Way & 0.2/0.8 & $110-140$/185 & spherical & Meaburn et al.\ (1996) \\
AG Carinae  & Milky Way & 0.87\,$\times$\,1.16 & 70 & bipolar ? &
Nota et al.\ (1992) \\
WRA 751 & Milky Way & 0.5 & 26 & bipolar & Weis (2000) \\
He 3-519 & Milky Way & 2.1 & 61 & spherical & Smith et al.\ (1994) \\
HD 168625 & Milky Way & 0.13\,$\times$\,0.17& 40 & spherical ? & 
Nota et al.\ (1996b)\\
Pistol Star & Milky Way & 0.8\,$\times$\,1.2 & 60 & spherical & Figer et al.\ (1999) \\
R127 & LMC & 1.3 & 32 & bipolar & this work \\
R143 & LMC & 1.2 & 24 (split)  & irregular & this work \\
S61 & LMC & 0.82 & 27 & spherical & this work \\
S Dor & LMC & $< 0.25$? & $< 40$ (FWHM)& ?  & this work \\
R\,71 & LMC & $< 0.1$? & 20  & ? & Stahl \& Wolf (1986b) \\
R\,99 & LMC & background & background & --- & this work\\
R\,84 & LMC & $< 0.3$ ? & 24 (split) & ? & this work \\
S119 & LMC ? & 1.8 & 26 & spherical/outflow & Weis et al.\ (2003) \\
Sk$-69\degr$279 & LMC & 4.5 & 14 & spherical/outflow & Weis \& Duschl (2002) \\
\hline
\hline
\end{tabular}
\end{center}
\end{table*}

\subsection{Expansion velocities}

\begin{figure}
\resizebox{\hsize}{!}{\includegraphics{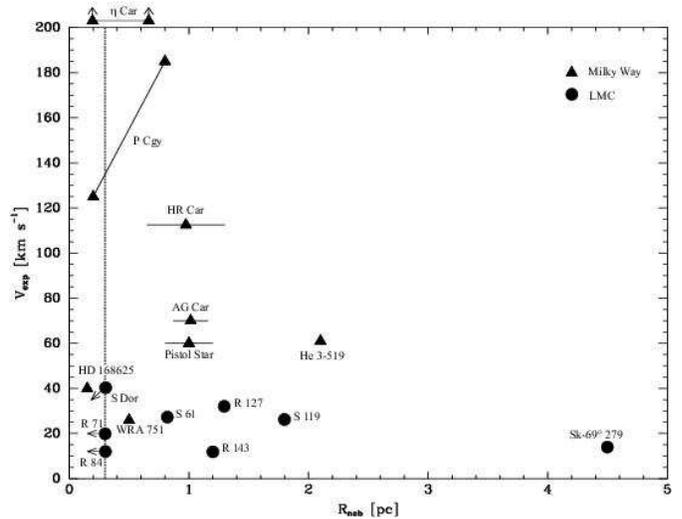}} \caption{
In this plot we compare the expansion velocities and sizes of nebulae around
LBVs in the Milky Way (triangles) and the LMC (circles). A dashed line
indicates the limit (0.3\,pc) up to which we can resolve nebulae in the LMC. 
In the case of asymetric nebular lines
show the limits in each symetry axis or connect datapoints of the same object
in case two nebula parts exists (P Cyg and $\eta$ Car). 
} \label{fig:rvexp} 
\end{figure}

While the sizes are about the same for  LBV nebulae, the 
expansion velocities are significantly different. 
Looking at nebulae in our Galaxy, it seems at first glance 
the larger a nebula, the lower its expansion 
(Fig. \ref{fig:rvexp}). This can manifest some kind
of evolution---in connection to a slow-down process---of LBV nebulae. 
As argued in Weis et al.\ (1997a) and Weis (2001), HR Car does show signs 
of being an aged, slowed down version of $\eta$ Car's nebula. We would 
expect that larger nebulae are older since they had more time to
expand, but at the same time the expansion will have
slowed them down. However, this scenario of larger nebulae being 
slower completely breaks down for the nebulae in the LMC. First,
we note that all
LBV nebulae in the LMC  are expanding  much more slowly than those in the 
Milky Way. In the LMC none is moving faster than 32\,\kms---in the Milky Way 
the expansion velocities are up to an order of magnitudes higher. 
The slowest LBV nebula in the Milky Way is about as fast as the fastest LMC 
nebula. As a
consequence,  the spread in expansion velocity of nebulae in the LMC is much
smaller than in the Galaxy. While we find velocities between 26\,\kms\ and
several 1000\,\kms\ in the Galaxy, the LMC LBVs range from only 12 to 
32\,\kms. The expansion velocities of LBV nebulae in the LMC are  
confined to a small range. 
From the presently known sample we conclude that there is a significant
difference in the expansion velocities between the nebulae in the 
LMC and the Milky Way. 
Even if we  exclude $\eta$ Car at that point, given its recent outburst
and exceptional status, this statement holds true.
Note in this context, that detecting higher expansion velocities in
nebulae around LBVs in the LMC is not a problem. Even if we are not resolving
the nebulae, the high expansion velocity will be visible in spectra which 
include the [N\,{\sc ii}], which will be broadend, splitted or at least
be shifted against the stellar radial velocity. 

So what determines the expansion velocities of LBV nebulae, and why are they so
different between the LMC and the Milky Way?
A lot of different stellar and environmental processes possibly
affect the expansion velocities, only the most obvious of which we 
will discuss in the following. Most likely, however, a 
mixture of the various processes will occur.
\begin{itemize}
\item{ {\it Stellar winds:} The mass loss history and change of 
wind velocity during the LBV phase is a function of the star's
initial mass, for instance. The formation of an LBV nebula is thought to occur 
due to wind-wind interaction, as proposed in, e.g., Garc{\'\i}a-Segura et 
al.\ (1996). If,
now, the input for wind velocity and mass loss is different, 
the expansion velocity of the nebulae will naturally differ
due to the solution of the wind-wind interaction models.    
The expansion velocities will also change, if the  duration
of the LBV phase 
(or more precisely the phase of slow wind)  is different, since then 
the interaction time of fast and slow wind will also change.}
Since we suspect that stellar wind influences the
    expansion velocities of the LBV nebulae, all effects that determine 
the wind velocity will automatically also influence our expansion velocities. 
In this context we will only briefly discuss one process that changes the
stellar wind velocity, namely metallicity or more general abundances,
since e.g. rotation might change the abundance in the outer layers of the
stars. If we have stars with lower metallicity/abundances---such 
as in the LMC 
objects---they generally have slower wind velocities, at least in their
main-sequence phase. Nothing is known so far about the wind velocity 
of stars in the LBV phase with lower metallicity/abundances. 
Therefore, different abundances---initially or, e.g., due to mixing from 
rotation---will also influence the expansion velocities.
\item{{\it Eruptions:} For some LBVs (e.g. $\eta$ Car, P Cyg) we know
    that they underwent a giant eruption, and at least parts of the
    nebulae were created in these outbursts. Whether or not all
 LBVs had such an outburst
    is not known (due to the lack of historical records). Therefore, the
    expansion velocities of the nebulae might be different since some 
resulted from outburst while outer nebulae formed through interaction. 
If we  assume that all LBV nebulae were created in outbursts, then
their outbursts can easily have a different strength and duration, which
would also result in a
variety of expansion velocities. Since we do not know what the underlying
mechanism of the giant eruption is, this process could be highly dependent on 
metallicity and then explain why LMC LBVs expand so much more slowly.}  
\item{{\it Environment:} Whatever created the nebulae, wind or eruption, the 
surrounding ISM will also have an  impact on the expanding nebulae. 
Density and temperature of the environment that harbors the LBVs will
have a noticeable influence on the nebulae. By sweeping up ISM, 
the nebulae will slow down  and start mixing. Instabilities will occur 
(e.g., Rayleigh-Taylor, Kelvin-Helmholtz) that can lead as far as
to the final disruption of the nebula and possible outflows as seen in, e.g.,
S\,119. So far, the closer environment in which the LBVs are situated 
has barely been studied, so that nothing is known about its 
influence on the LBV nebulae.  }
\end{itemize}
Some of the above mentioned parameters---perhaps all---are responsible
for the different expansion velocities of the LBVs in the LMC and the 
Milky Way. As a natural consequence  a comparison of the nebula with 
stellar parameters would be made. For a specific reason this comparison has
explicitly been awoided. LBVs do change their stellar parameters
significantly in timescales as short as some years. Therefore, a simple 
comparison of stellar parameters of LBVs with each other and with the 
nebula parameters
would lead to invalid conclusions. The stellar parameters, especially the wind
velocity, were different at the time the nebulae formed. Comparing the 
stellar wind, as detected in the current spectra, would not help 
to pin down the different properties of the nebulae in the LMC and the 
Milky Way.

LBV nebulae in the LMC are similar in shape and size 
to those in the Milky Way, but their expansion velocities are different. 
Nebulae are either formed in giant eruptions or due to wind-wind 
interaction. We conclude that the formation of
LBV nebulae seem similar for each nebula, but must occur with different 
strengths to account for the different velocities. At least to some degree, 
metallicity seems to play a role in the formation of
LBV nebulae as the different expansion velocities of LMC and Galactic nebulae 
indicate.

\begin{acknowledgements}
Special thanks go to Dominik J.\ Bomans (Bochum) for his great help and 
endless efforts in this work and his supporting discussions on the subject.
I am grateful to Wolfgang J.\ Duschl (Heidelberg) who help considerably to
improve the manuscript and to Roberta
M.\ Humphreys (Minnesota), and Kris Davidson (Minnesota), who 
supported this work with their input.
I am obliged to Otmar Stahl for proving independent 
radial velocity measurements for R\,143.
I thank the referee for helpful comments and suggestions.

Partly based on observations made with the European
Southern Observatory telescopes obtained from the ESO/ST-ECF Science
Archive Facility.
Based partly on observations made with the NASA/ESA Hubble Space Telescope,
obtained from the data archive at the Space Telescope Institute. STScI is
operated by the association of Universities for Research in Astronomy,
Inc. under the NASA contract  NAS 5-26555.
This work was partially supported by the DFG through grant Du\,168/8-1. 
The data reduction and analysis was in part carried out on a workstation 
provided by the {\it Alfried Krupp von Bohlen und Halbach Stiftung\/}. 

\end{acknowledgements}

\end{document}